\documentclass[aps,prl,preprint,superscriptaddress]{revtex4-2}

\draft 
\usepackage{graphicx}
\usepackage{dcolumn}
\usepackage{bm}

\usepackage[utf8]{inputenc}
\usepackage[T1]{fontenc}
\usepackage{mathptmx}
\usepackage{xcolor}
\usepackage{units}
\begin{document}


\title{A unipolar quantum dot diode structure for advanced quantum light sources} 



\author{T. Strobel}
\thanks{These authors contributed equally}
\affiliation{Institut f\"ur Halbleiteroptik und Funktionelle Grenzfl\"achen, Center for Integrated Quantum Science and Technology ($IQ^{ST}$) and SCoPE, University of Stuttgart, Allmandring 3, 70569 Stuttgart, Germany}

\author{J. H. Weber}
\thanks{These authors contributed equally}
\affiliation{Institut f\"ur Halbleiteroptik und Funktionelle Grenzfl\"achen, Center for Integrated Quantum Science and Technology ($IQ^{ST}$) and SCoPE, University of Stuttgart, Allmandring 3, 70569 Stuttgart, Germany}

\author{M. Schmidt}
\affiliation{Lehrstuhl für Angewandte Festkörperphysik, Ruhr-Universität Bochum, D-44780 Bochum, Germany}

\author{L. Wagner}
\affiliation{Institut f\"ur Halbleiteroptik und Funktionelle Grenzfl\"achen, Center for Integrated Quantum Science and Technology ($IQ^{ST}$) and SCoPE, University of Stuttgart, Allmandring 3, 70569 Stuttgart, Germany}

\author{L. Engel}
\affiliation{Institut f\"ur Halbleiteroptik und Funktionelle Grenzfl\"achen, Center for Integrated Quantum Science and Technology ($IQ^{ST}$) and SCoPE, University of Stuttgart, Allmandring 3, 70569 Stuttgart, Germany}

\author{M. Jetter}
\affiliation{Institut f\"ur Halbleiteroptik und Funktionelle Grenzfl\"achen, Center for Integrated Quantum Science and Technology ($IQ^{ST}$) and SCoPE, University of Stuttgart, Allmandring 3, 70569 Stuttgart, Germany}

\author{A. D. Wieck}
\affiliation{Lehrstuhl für Angewandte Festkörperphysik, Ruhr-Universität Bochum, D-44780 Bochum, Germany}

\author{S. L. Portalupi}
\affiliation{Institut f\"ur Halbleiteroptik und Funktionelle Grenzfl\"achen, Center for Integrated Quantum Science and Technology ($IQ^{ST}$) and SCoPE, University of Stuttgart, Allmandring 3, 70569 Stuttgart, Germany}

\author{A. Ludwig}
\affiliation{Lehrstuhl für Angewandte Festkörperphysik, Ruhr-Universität Bochum, D-44780 Bochum, Germany}

\author{P. Michler}
\affiliation{Institut f\"ur Halbleiteroptik und Funktionelle Grenzfl\"achen, Center for Integrated Quantum Science and Technology ($IQ^{ST}$) and SCoPE, University of Stuttgart, Allmandring 3, 70569 Stuttgart, Germany}



\date{\today}

\begin{abstract}
Triggered, indistinguishable, single photons play a central role in various quantum photonic implementations. Here, we realize a novel n$^+-$i$-$n$^{++}$ diode structure embedding semiconductor quantum dots: the gated device enables spectral tuning of the transitions and deterministic control of the observed charged states. Blinking-free single-photon emission and high two-photon indistinguishability is observed. The linewidth's temporal evolution is investigated for timescales spanning more than $6$~orders of magnitude, combining photon-correlation Fourier spectroscopy, high-resolution photoluminescence spectroscopy, and two-photon interference (visibility of $V_{\text{TPI,}\unit[2]{ns}}=\unit[\left(85.5\pm2.2\right)]{\%}$ and $V_{\text{TPI,}\unit[9]{ns}}=\unit[\left(78.3\pm3.0\right)]{\%}$). No spectral diffusion or decoherence on timescales above $\sim\unit[9]{ns}$ is observed for most of the dots, and the emitted photons' linewidth ($\unit[\left(420\pm30\right)]{MHz}$) deviates from the Fourier-transform limit only by a factor of $\unit[1.68]{}$. Thus, for remote TPI experiments, visibilities above $74\%$ are anticipated. The presence of n-doping only signifies higher available carrier mobility, making the presented device highly attractive for future development of high-speed tunable, high-performance quantum light sources.

\end{abstract}


\maketitle 

Quantum optical implementations and applications require sources of non-classical light capable of emitting single photons on-demand and with a high degree of indistinguishability~\cite{Michler_2017}. Furthermore, for upscaling the experimental complexity, remote sources capable of emitting light at the same wavelength are highly desirable~\cite{Patel2010b, Giesz2015a, Weber2018c, Weber2019b, Zhai2022a}. Semiconductor quantum dots have shown potentials to fulfill all the aforementioned needs, additionally being able to control the emission wavelength via various mechanisms, i.e. the application of electric~\cite{Patel2010b} or magnetic field~\cite{Akopian2010}, as well as the use of external mechanical strain~\cite{Trotta2012, Martin-Sanchez2018}.
\noindent Being embedded in a semiconductor matrix allows for the realization of high-quality photonic resonators~\cite{Wang2016a, Wang2019h, Liu2019b}, as well as the implementation of compact diode structures to control the emission wavelength via local  tuning~\cite{Somaschi2016a}. Interestingly, placing the quantum dots (QDs) into an electrically gated structure further provides a stabilization of the carrier environment which decreases the impact of spectral diffusion on the photon linewidth~\cite{Kuhlmann2015a,Prechtel2016a}. These possibilities culminated in the realization of a source of single- and indistinguishable-photons with an end-to-end efficiency as high as 57\%~\cite{Tomm2021a}, having the sample been grown via molecular beam epitaxy (MBE) embedding the emitters into a p-i-n diode structure. Despite these recent results, further improvements in the sample design and realization can be obtained: from the growth point of view, the presence of a p-layer in the implementation of the diode embedding the QDs requires the doping with carbon atoms which constitute an impurity during growth as well as in the MBE chamber itself, even though there is a minimal memory effect~\cite{Reuter1999,Prechtel2016a,Ludwig2017}. In high-frequency electrical device applications, using holes in p-doped structures instead of electrons in n-doped structures would limit the achievable speed due to their $20$-times lower mobility allowing for fast control~\cite{Pedersen2020a}. For example, a scheme theoretically proposed in Ref.~\cite{Bauch2021b} indeed requires electrical Stark shift of the QD transitions on sub-picosecond timescales to implement cavity-mediated single and photon-pair generation with QDs.\\
For these reasons, in the present study, the quantum dots have been embedded into a n$^+-$i$-$n$^{++}$ diode structure, while the light extraction is enhanced by a planar cavity design formed by two distributed Bragg reflectors (DBR) above and below the emitters. This novel diode structure was implemented performing a two-step growth combining metal organic vapor phase epitaxy (MOVPE) for the n-doped, bottom DBR, and MBE for the QD and gate diode, as well as the top DBR. On the one hand, MOVPE allows for a fast deposition of high-quality multilayers forming high-reflectivity DBRs, being the growth performed at high temperature and low vacuum. On the other hand, MBE with its high-vacuum and high purity growth conditions, enables the slow and controlled deposition of semiconductor nanostructures with a defect-free environment, resulting in high-coherence photon emission~\cite{Mccray2007,Ludwig2017,Najer2019b,Zhai2022a,Kosarev2022}. In the following, the optical and quantum optical properties of the emitted photons from the n$^+-$i$-$n$^{++}$ diode structure will be reported. In particular, high single-photon purity and indistinguishability have been observed respectively via Hanbury-Brown and Twiss, and Hong-Ou-Mandel-type experiments. Fabry-Perot interferometry allowed for measuring the static spectra of the emitted photons in high resolution and for comparing the linewidth with the expected Fourier-transform (FT) limit (obtained from decay time measurements).  Photon-correlation Fourier spectroscopy (PCFS) ~\cite{Brokmann2006b,Schimpf2019b} proves that spectral broadening mechanisms are absent in the large majority of the investigated QDs, while small deviations from the FT limit act on timescales shorter than a few nanoseconds, in accordance with two-photon interference measurements. In line with pioneering p-i-n-type diodes the novel n$^+-$i$-$n$^{++}$ diode structure further enables the Stark tuning of the QD transitions, together with a precise control of the observed charge states in the dot. Finally, voltage dependent studies prove the stabilization effect of the diode structure, having a drastic reduction of the emission linewidth for the designed operation conditions.

\section{Device design and resonant excitation}
\begin{figure*}
	\centering
	\includegraphics[width=1\linewidth]{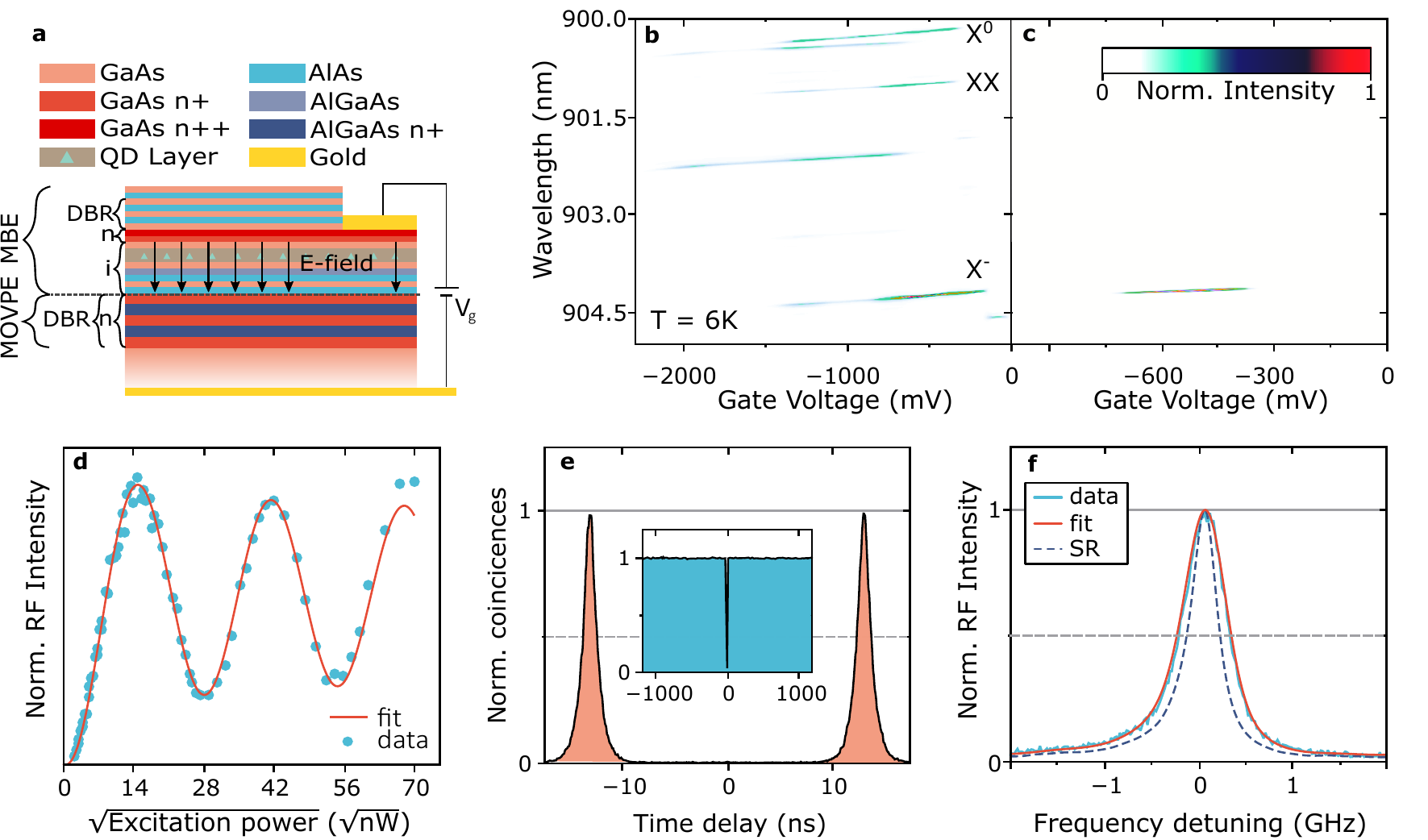}
	\caption{\textbf{Sample description and characterization:} \textbf{a}, Schematic of the sample structure. \textbf{b}, $\mu$-PL of one selected quantum dot as a function of gate voltage using an AB pumping scheme. \textbf{c}, RF signal of the same QD, under resonant excitation of the trion transition. \textbf{d}, Resonant fluorescence intensity over the square root of the (pulsed) laser power. \textbf{e}, Second-order autocorrelation measurement of the investigated trion line: in the inset, long time delay data (with \unit[13]{ns} binning) show the absence of blinking. The data are normalized to the Poissonian level. \textbf{f}, High-resolution FPI measurement of the trion linewidth (under pulsed resonant excitation for $V_{\text{g}}=\unit[-540]{mV}$ applied voltage). Data (solid blue), fit (solid red) and the system response (SR) function (dashed) are shown.}
	\label{fig:fig1}
\end{figure*}
The sample employed has been grown by combining MOVPE and MBE techniques. A schematic view of its structure is depicted in Fig.~\ref{fig:fig1}a. In a first step, n$^{+}$-doped bottom DBRs are grown with MOVPE. Then, the sample is removed from the reactor and shipped for the subsequent MBE covering of the InAs QDs embedded in the n$^+-$i$-$n$^{++}$ gate diode (see Methods).\\  

We first investigate the micro-photoluminescence ($\mu$-PL) spectrum (in above barrier (AB) pumping scheme at saturation) as a function of the applied gate voltage $V_{\text{g}}$. The acquired data are depicted as a voltage/wavelength intensity map in Fig.~\ref{fig:fig1}b. Starting from a negative bias, several distinct lines appear and disappear as $V_{\text{g}}$ is increased towards \unit[0]{V}. Two-photon excitation (TPE) and decay time measurements (see Supplementary Fig.~S1 and S3), allow for identifying the transitions as neutral exciton and biexciton, and negatively charged trion respectively (as this transition appears at a bias where a negative charged QD is energetically more favorable). At low pumping rates (see Supplementary Fig.~S2), only trion or exciton lines are present for specific applied voltage values~\cite{Lobl2017a}: this controlled switching can be understood as a consequence of the Coulomb blockade. By applying gate voltages, the position of the discrete energy levels in the QD relative to the Fermi sea changes. Depending on the voltage level certain excitonic states can be favored.\\

The emission lines undergo a wavelength shift (\unit[$71.1\pm0.2$]{GHz~V$^{-1}$}, $\sim$\unit[$2.8\cdot10^{-3}$]{GHz~per~V~cm$^{-1}$}) with changing $V_{\text{g}}$. This change in wavelength is the characteristic dc Stark shift~\cite{Wen1995b} caused by the electrostatic field of the diode structure. These findings demonstrate the controlled switching of electronic states in this novel structure as well as the wavelength tunability with applied voltage.\\

Fig.~\ref{fig:fig1}c shows the evolution of the trion wavelength under resonant pumping for various applied voltages, where again a Stark shift of the emission wavelength can be observed (\unit[$72.7\pm0.6$]{GHz~V$^{-1}$}, $\sim$\unit[$2.9\cdot10^{-3}$]{GHz~per~V~cm$^{-1}$}). Under pulsed excitation, the trion resonance fluorescence intensity shows clear Rabi oscillations (see Fig.~\ref{fig:fig1}d), proving coherent control of the state population: the observed oscillations yield a state preparation fidelity of $\approx\unit[85]{\%}$. 



Under resonant excitation at the $\pi$-pulse, the second-order correlation plotted in Fig.~\ref{fig:fig1}e reveals a pronounced anti-bunching at zero time delay, with a single-photon purity of $g^{\left(2\right)}\left(0\right)=0.028\pm0.001$. Furthermore no blinking is observed under resonant pumping in the whole time delays considered in the histogram of correlations (as shown in the inset of Fig.~\ref{fig:fig1}e with time separations beyond $\pm\unit[1]{\mu s}$): this represents an important characteristic for the implementation of high brightness sources of quantum light. Besides that, the absence of p-doping helps in decreasing the presence of impurities in the QD host material, impurities which may also result in blinking dynamics at long time separation~\cite{Houel2014}. This blinking-free behavior is consistent with studies on similar diode-like heterostructures utilizing the Coulomb blockade to stabilize the charge environment~\cite{Zhai2020a}. For different applied voltages, it is also shown that resonant two-photon excitation (TPE) can be employed to generate exciton-biexciton pairs (see Supplementary Fig.~S1).\\

A charge-stable QD environment, in combination with a low density of impurities and defects, is also supposed to positively impact the emission linewidth. For this reason, we investigate the emission linewidth under resonant pumping at the $\pi$-pulse. First, a scanning Fabry-Perot interferometer (FPI) is employed for recording high-resolution stationary spectra of the investigated QDs (see Methods). Fig.~\ref{fig:fig1}f shows the measured emitted photons' linewidth for the trion transition under investigation reaching a value of $\unit[\left(420\pm30\right)]{MHz}$, broadened only by a factor of 1.68 compared to the FT limit ($\Delta\nu_{\text{FL}}=\unit[\left(250\pm20\right)]{MHz}$). This measured close-to lifetime-limited linewidth demonstrates a modest impact of dephasing and spectral diffusion in the presented heterostructure.\\

\section{Linewidth temporal evolution over applied gate voltage}
\begin{figure*}
	\centering
	\includegraphics[width=1\linewidth]{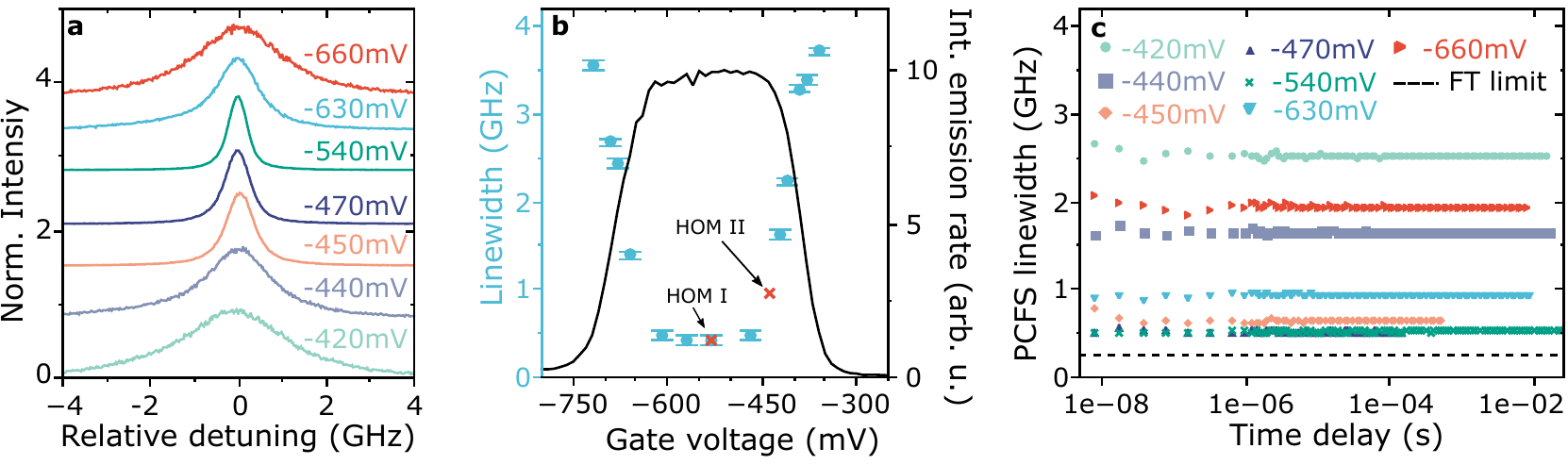}
	\caption{\textbf{Linewidth investigation of the trion line at the $\pi$-pulse for different gate voltage levels:} \textbf{a}, High resolution FPI spectra for different gate voltages (vertically shifted for clarity). \textbf{b}, Emission linewidth extracted from the FPI data in \textbf{a} (blue dots). For comparison the RF emission intensity extracted from the voltage map in Fig.~\ref{fig:fig1}\textbf{c} is shown in black. The two crosses indicate the conditions for the successive two-photon interference measurements of Fig.~\ref{fig:fig3} and Tab.~\ref{table:1}. \textbf{c}, PCFS measurements from ns to ms timescales. The estimated lifetime-limited linewidth is shown in black (FL). For each voltage the linewidth is a straight line indicating no spectral dynamics on the mentioned timescales. Moving away from the center position of the charge plateau in the voltage scan an increase of the emission linewidth is observed.}
	\label{fig:fig2}
\end{figure*}

Following this first characterization, we conduct FPI measurements for various applied gate voltages, and the results are summarized in Fig.~\ref{fig:fig2}a. As it can be seen, the observed linewidth varies drastically with the applied bias. The emission linewidth extracted from fitting the FPI measurements is reported in Fig.~\ref{fig:fig2}b (blue dots): interestingly, a voltage range exists for which the linewidth reaches minimal values which corresponds to a maximum in the observed $\mu$-PL intensity (black solid line). Outside this plateau, the measured linewidth increases, accompanied by a decrease in luminescence. In line with previous studies, the observed line broadening can be attributed to cotunneling~\cite{Smith2005,Dreiser2008,Latta2009,Reigue2018a}: the QD is tunnel-coupled to the Fermi sea, with increased tunneling interaction for voltage levels outside the shown stable region, hence decreasing the photon coherence. At the center of the plateau tunnel coupling is inhibited increasing the photon coherence time, hence minimizing the linewidth.

Despite the narrow linewidth observed at the center of the voltage plateau, it is useful to investigate the origin of potential dephasing and spectral diffusion mechanisms responsible for the small deviation from the Fourier limit. To do so, PCFS is employed~\cite{Brokmann2006b,Schimpf2019b}. PCFS enables measuring the time evolution of the linewidth of the emitted photons, with high temporal and spectral resolution and for timescales that can vary from few nanoseconds to few milliseconds (see Methods).\\
Exemplary results of PCFS measurements are reported in Fig.~\ref{fig:fig2}c, where the applied gate voltage has also been varied in order to provide a profound insight on the impact of the applied electric field on the emission linewidth. As it can be seen, the PCFS results show that, for all applied voltages, the linewidth remains constant over time, where only the absolute value is changing. The lowest measured linewidth in PCFS is $\unit[520\pm100]{MHz}$ ($V_{\text{g}}=\unit[-570]{mV}$) which is very close to the FT limit (dashed line) as it also was observed in the previous FPI measurement recorded for similar applied voltages (see Fig.~\ref{fig:fig1}f). The PCFS measured linewidth is close to the resolution limit of the PCFS setup, which could explain the slightly higher value with respect to the analogous FPI value, which corresponds to the stationary limit of the emission spectrum~\cite{Vural2020a}. This agreement between PCFS at long timescales and FPI measurements indicate that any deviation from the FT limit is due to dephasing and spectral diffusion mechanisms happening at timescales shorter than \unit[10]{ns}. The same conclusion applies for the other applied voltages, since the long timescale PCFS values match the corresponding FPI measurements (see Fig.~\ref{fig:fig2}b and Supplementary Fig.~S4). This supports the conclusion that the process responsible for this additional line broadening happens at a timescale shorter than $\unit[10]{ns}$, since there is no further broadening, even for times longer than $\unit[10]{ms}$.

The discussed results, further supported by previous studies~\cite{Smith2005,Kuhlmann2015a,Reigue2018a}, clearly indicate the existence of a voltage range for which the linewidth is minimized. There, the small deviation from FT limit is proven to be due to mechanisms with dynamics faster than $\unit[10]{ns}$.

\section{Short timescale dynamics via voltage-dependent two-photon interference}

\begin{figure}
	\centering
	\includegraphics[width=1\linewidth]{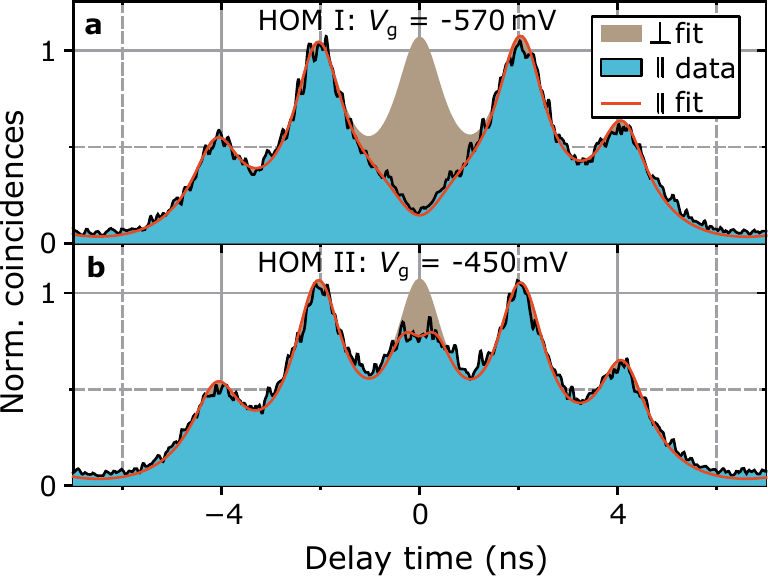}
	\caption{\textbf{Two-photon interference:} TPI measurement from the trion at the $\pi$-pulse for two different gate voltages. Double pulses with \unit[2]{ns} temporal separation were used to create the interfering photons. Unwanted laser background (or broad tails from phonon assisted emission) is filtered via a transmission spectrometer of spectral width $\Delta_{\text{filter}} = \unit[15]{GHz}$, much larger than the emission linewidth. The blue areas show the measured data for parallel polarization, with their fit function in red. The brown areas correspond to the fitted data of the orthogonal polarization measurement. \textbf{a}, TPI results for a voltage of $V_{\text{g}}=\unit[-570]{mV}$ corresponding to the center of the charge plateau. \textbf{b}, measurement results for a voltage of $V_{\text{g}}=\unit[-450]{mV}$ at the edge of the charge plateau (compare with Fig.~\ref{fig:fig2}\textbf{b}).}
	\label{fig:fig3}
\end{figure}

\noindent To probe even shorter timescales, two-photon inference (TPI) measurements with consecutively emitted photons are conducted (time separation between $\unit[2]{ns}$ and $\unit[9]{ns}$), therefore extending the investigated time range close to the radiative lifetime of the emitter. The measurements in Fig.~\ref{fig:fig3}a show the central peaks around zero time delay with a clear signature of two-photon interference, for an applied gate voltage within the previously observed plateau (here $V_{\text{g}}=\unit[-570]{mV}$). In this configuration, a TPI visibility as high as $V_{\text{TPI}}^{\unit[-570]{mV}}=\unit[\left(85.8\pm2.2\right)]{\%}$ is recorded for a photon time separation of $\unit[2]{ns}$.
This implies that the mechanism responsible for the small deviation from near-unity visibility happens at timescales shorter than $\unit[2]{ns}$. Considering the rather small size of the QDs, the effective coupling to acoustic phonons~\cite{Zibik2009} could result in a dephasing mechanism which acts at below nanosecond timescales: this would explain the deviation from FT limit and the respective impact on the TPI.\\
To further investigate the broadening mechanisms, TPI measurements with time separations of $\unit[4]{ns}$ and $\unit[9]{ns}$ were performed. An overview of the results is given in Tab.~\ref{table:1}. For the center of the charge plateau at $V_{\text{g}}=\unit[-570]{mV}$ a decrease from $V_{\text{TPI,} \unit[2]{ns}} = \unit[\left(85.5\pm2.2\right)]{\%}$ to $ V_{\text{TPI,} \unit[4]{ns}} = \unit[\left(79.7\pm2.5\right)]{\%}$ (for $\unit[4]{ns}$ time separation) and to $ V_{\text{TPI,} \unit[9]{ns}} = \unit[78.3\pm3.0]{\%}$ (for $\unit[9]{ns}$) in the TPI visibility is observed. This rather modest decrease of the TPI visibility can be attributed to spectral broadening mechanisms happening at such timescales (still not resolvable in PCFS measurements). Intriguingly, estimating the reachable TPI visibility from the recorded stationary linewidth in FPI measurements (Fig.~\ref{fig:fig2}a), a value of $V_{\text{TPI}}^{\text{sim}} = \unit[\left(74.3\pm1.5\right)]{\%}$ is expected~\cite{Kambs2018b}.  This quantity would also represent the achievable TPI visibility for photons stemming from remote sources. This value matches, within margin of uncertainty, the observed value of $V_{\text{TPI,} \unit[9]{ns}} = \unit[\left(78.3\pm3.0\right)]{\%}$. This indicates that, together with broadening mechanisms happening on timescales smaller than $\unit[2]{ns}$, a small linewidth broadening is further observed between $\unit[2]{ns}$ and $\unit[9]{ns}$. Above this timescale value, no further broadening mechanisms are observed: this is confirmed independently by PCFS measurements, and by the agreement between the observed TPI at $\unit[9]{ns}$ and the expected two-photon interference inferred from the stationary spectrum (measured with FPI measurements).

Interestingly, for voltages outside the discussed stable region (here $V_{\text{g}}=\unit[-450]{mV}$), the observed TPI is drastically reduced. As shown in Fig.~\ref{fig:fig3}b, a value as low as $V_{\text{TPI}}^{\unit[-450]{mV}}=\unit[\left(14.9\pm3.6\right)]{\%}$ is observed. The applied voltage has been chosen in a range which has already a clear impact on the emission linewidth, despite a modest reduction of the observed count rate (see Fig.~\ref{fig:fig2}b). These findings are consistent with the observation on the line broadening: while for the voltage range within the plateau the electric field stabilizes the environment, resulting in a close to FT limited linewidth, outside this charge plateau the situation changes. There, carrier tunneling induces a broadening of the linewidth, and the tunneling itself happens on time scales shorter than $\unit[10]{ns}$ (as seen in PCFS data) and than $\unit[2]{ns}$ (as seen in TPI measurements). More than \unit[95]{\%} of the investigated quantum dots (20 QDs) showed a linewidth behavior comparable with the reported results (As depicted in Supplementary Fig.~S5, only one dot showed linewidth temporal dynamics in PCFS data, compatible with the presence of a local carrier trap next to the dot). Outside this stable plateau, the TPI visibility remains low ($\approx 14\%$) for all photon time separations and it matches the value inferred from the linewidth: once again, this confirms the impact of dephasing on timescales shorter than $\unit[2]{ns}$ (see results in Table~\ref{table:1}).\\

\begin{table}
\begin{tabular}{c|cccc}
	\hline\hline
	V$_{\text{g}}$(mV)& V$_{\text{TPI,\unit[2]{ns}}}$(\%) & V$_{\text{TPI,\unit[4]{ns}}}$(\%) & V$_{\text{TPI,\unit[9]{ns}}}$(\%) & V$^{\text{sim}}_{\text{TPI}}$(\%) \\ 
	\hline 
	$-570$ & $85.5\pm2.2$ & $79.7\pm2.5$ & $78.3\pm3.0$ & $74.3\pm1.5$ \\ 
	
	$-450$ & $14.9\pm3.6$ & $13.9\pm3.7$ & $14.6\pm3.5$ & $14.5\pm1.0$  \\ 
	\hline\hline
\end{tabular}
\caption{TPI visibilities for different photon temporal delays and applied gate voltages. Moreover the expected TPI visibility (V$^{\text{sim}}_{\text{TPI}}$) of statistically independent QD emissions is extracted from the static FPI measurements.}
\label{table:1} 
\end{table}


\section{Conclusions}
In this study, MOVPE and MBE have been combined to realize high optical quality quantum dots, embedded into a novel n$^+-$i$-$n$^{++}$ diode, and within a planar cavity formed by two DBRs. The high quality DBR is ensured by the high temperature (and high speed) deposition of MOVPE, while the MBE ensures the growth of QDs within low defect material structure. This, in combination with the embedding diode enables the observation of close-to-FT limit QD linewidth ($\unit[\left(420\pm30\right)]{MHz}$), where high coherent control of the population is observed as well as high single-photon purity ($g^{\left(2\right)}\left(0\right)=0.028\pm0.001$). The sample is investigated combining PCFS, FPI and TPI measurements enabling a study on emission linewidth and dynamics of the decoherence mechanisms for timescales spanning more than $6$ orders of magnitude. In particular, the small deviation from lifetime-limited spectra is investigated via PCFS and FPI measurements: combining these two techniques it is possible to conclude that no spectral broadening mechanisms happen for timescales between $\unit[10]{ns}$ and the stationary limit (set by the FPI results). Furthermore, it has been clearly proven that the applied gate voltage allows for wavelength tuning and stabilizing the QD charge environment, within a defined voltage range where the luminescence is maximized. Outside from this ideal plateau, the emission linewidth increases and the brightness decreases. Still, thanks to PCFS we can conclude that the mechanisms responsible for this degradation outside the ideal voltage range have a timescale below $\unit[10]{ns}$. Finally, two-photon interference measurements are employed to investigate the timescales between $\unit[2]{ns}$ and $\unit[9]{ns}$ to provide further insights in a time range not accessible by PCFS. These measurements show that, even for the smallest time separation, the linewidth is still slightly impacted by fast dephasing mechanisms (below $\unit[2]{ns}$), which we attribute to acoustic phonon coupling. Still, a small degradation of the TPI visibility is observed increasing the photon time separation to $\unit[9]{ns}$: in this condition, the measured visibility is consistent with the one expected by simulations, taking into account the stationary linewidth measured via FPI. This further demonstrates that the only decoherence mechanisms, despite modest, are happening at relatively short timescales. 

Phonon coupling can be reduced in the future by growing larger QDs (which couple less to phonons) or by integrating the emitters in photonic microcavities: Purcell enhancement is known to be beneficial for decreasing the impact of dephasing mechanisms, in particular at short timescales, further improving the photon indistinguishability. Both approaches, i.e. growth of larger QDs and use of photonic cavities, are fully compatible with the realized sample design. Interestingly, the absence of p-doping will enable in the near future the realization of devices where a fast AC bias can be applied, thanks to the high electron mobility. Finally, the observed Stark shift induced by the diode structure will have important application in realizing wavelength tunable sources, fundamental requirement for the implementation of TPI from distinct devices: already at the present stage, a remote visibility of $\approx 74\%$ would be expected, a value much higher than other reported for InGaAs QDs, only exceeded recently by GaAs quantum dots~\cite{Zhai2022a}. This makes the implementation of an n$^+-$i$-$n$^{++}$ diode, embedding (larger) QDs into optical microcavities highly desired for future quantum photonics.


\subsection{Acknowledgments}
The authors gratefully acknowledge the funding by the German Federal Ministry of Education and
Research (BMBF) via the project QR.X (No.~16KISQ013) and the company Quantum Design for their persistent support. We would like to thank Sergej Vollmer for the support in MOVPE growth.

\subsection{Author contributions}
T.S., J.H.W. and L.W. performed the measurements and analyzed the data. M.S. and A.L. grew the sample with the support of A.D.W.. L.E. fabricated the device. A.L. and M.J. designed the sample. T.S. and S.L.P. wrote the manuscript with the support of J.H.W and P.M.. A.L., M.J., S.L.P., A.D.W. and P.M. coordinated the project. All authors contributed to scientific discussions and revision of the manuscript.


%
%

%



\begin{thebibliography}{38}%
		\makeatletter
		\providecommand \@ifxundefined [1]{%
			\@ifx{#1\undefined}
		}%
		\providecommand \@ifnum [1]{%
			\ifnum #1\expandafter \@firstoftwo
			\else \expandafter \@secondoftwo
			\fi
		}%
		\providecommand \@ifx [1]{%
			\ifx #1\expandafter \@firstoftwo
			\else \expandafter \@secondoftwo
			\fi
		}%
		\providecommand \natexlab [1]{#1}%
		\providecommand \enquote  [1]{``#1''}%
		\providecommand \bibnamefont  [1]{#1}%
		\providecommand \bibfnamefont [1]{#1}%
		\providecommand \citenamefont [1]{#1}%
		\providecommand \href@noop [0]{\@secondoftwo}%
		\providecommand \href [0]{\begingroup \@sanitize@url \@href}%
		\providecommand \@href[1]{\@@startlink{#1}\@@href}%
		\providecommand \@@href[1]{\endgroup#1\@@endlink}%
		\providecommand \@sanitize@url [0]{\catcode `\\12\catcode `\$12\catcode
			`\&12\catcode `\#12\catcode `\^12\catcode `\_12\catcode `\%12\relax}%
		\providecommand \@@startlink[1]{}%
		\providecommand \@@endlink[0]{}%
		\providecommand \url  [0]{\begingroup\@sanitize@url \@url }%
		\providecommand \@url [1]{\endgroup\@href {#1}{\urlprefix }}%
		\providecommand \urlprefix  [0]{URL }%
		\providecommand \Eprint [0]{\href }%
		\providecommand \doibase [0]{https://doi.org/}%
		\providecommand \selectlanguage [0]{\@gobble}%
		\providecommand \bibinfo  [0]{\@secondoftwo}%
		\providecommand \bibfield  [0]{\@secondoftwo}%
		\providecommand \translation [1]{[#1]}%
		\providecommand \BibitemOpen [0]{}%
		\providecommand \bibitemStop [0]{}%
		\providecommand \bibitemNoStop [0]{.\EOS\space}%
		\providecommand \EOS [0]{\spacefactor3000\relax}%
		\providecommand \BibitemShut  [1]{\csname bibitem#1\endcsname}%
		\let\auto@bib@innerbib\@empty
		\bibitem [{\citenamefont {Michler~(Ed.)}(2017)}]{Michler_2017}%
		\BibitemOpen
		\bibfield  {author} {\bibinfo {author} {\bibfnamefont {P.}~\bibnamefont
				{Michler~(Ed.)}},\ }\href@noop {} {\emph {\bibinfo {title} {Quantum Dots for
					Quantum Information Technologies}}}\ (\bibinfo  {publisher} {Springer},\
		\bibinfo {year} {2017})\BibitemShut {NoStop}%
		\bibitem [{\citenamefont {Patel}\ \emph {et~al.}(2010)\citenamefont {Patel},
			\citenamefont {Bennett}, \citenamefont {Farrer}, \citenamefont {Nicoll},
			\citenamefont {Ritchie},\ and\ \citenamefont {Shields}}]{Patel2010b}%
		\BibitemOpen
		\bibfield  {author} {\bibinfo {author} {\bibfnamefont {R.~B.}\ \bibnamefont
				{Patel}}, \bibinfo {author} {\bibfnamefont {A.~J.}\ \bibnamefont {Bennett}},
			\bibinfo {author} {\bibfnamefont {I.}~\bibnamefont {Farrer}}, \bibinfo
			{author} {\bibfnamefont {C.~A.}\ \bibnamefont {Nicoll}}, \bibinfo {author}
			{\bibfnamefont {D.~A.}\ \bibnamefont {Ritchie}},\ and\ \bibinfo {author}
			{\bibfnamefont {A.~J.}\ \bibnamefont {Shields}},\ }\bibfield  {title}
		{\bibinfo {title} {{Two-photon interference of the emission from electrically
					tunable remote quantum dots}},\ }\href
		{https://doi.org/10.1038/nphoton.2010.161} {\bibfield  {journal} {\bibinfo
				{journal} {Nat. Photonics}\ }\textbf {\bibinfo {volume} {4}},\ \bibinfo
			{pages} {632} (\bibinfo {year} {2010})}\BibitemShut {NoStop}%
		\bibitem [{\citenamefont {Giesz}\ \emph {et~al.}(2015)\citenamefont {Giesz},
			\citenamefont {Portalupi}, \citenamefont {Grange}, \citenamefont
			{Ant{\'{o}}n}, \citenamefont {{De Santis}}, \citenamefont {Demory},
			\citenamefont {Somaschi}, \citenamefont {Sagnes}, \citenamefont
			{Lema{\^{i}}tre}, \citenamefont {Lanco}, \citenamefont {Auff{\`{e}}ves},\
			and\ \citenamefont {Senellart}}]{Giesz2015a}%
		\BibitemOpen
		\bibfield  {author} {\bibinfo {author} {\bibfnamefont {V.}~\bibnamefont
				{Giesz}}, \bibinfo {author} {\bibfnamefont {S.~L.}\ \bibnamefont
				{Portalupi}}, \bibinfo {author} {\bibfnamefont {T.}~\bibnamefont {Grange}},
			\bibinfo {author} {\bibfnamefont {C.}~\bibnamefont {Ant{\'{o}}n}}, \bibinfo
			{author} {\bibfnamefont {L.}~\bibnamefont {{De Santis}}}, \bibinfo {author}
			{\bibfnamefont {J.}~\bibnamefont {Demory}}, \bibinfo {author} {\bibfnamefont
				{N.}~\bibnamefont {Somaschi}}, \bibinfo {author} {\bibfnamefont
				{I.}~\bibnamefont {Sagnes}}, \bibinfo {author} {\bibfnamefont
				{A.}~\bibnamefont {Lema{\^{i}}tre}}, \bibinfo {author} {\bibfnamefont
				{L.}~\bibnamefont {Lanco}}, \bibinfo {author} {\bibfnamefont
				{A.}~\bibnamefont {Auff{\`{e}}ves}},\ and\ \bibinfo {author} {\bibfnamefont
				{P.}~\bibnamefont {Senellart}},\ }\bibfield  {title} {\bibinfo {title}
			{{Cavity-enhanced two-photon interference using remote quantum dot
					sources}},\ }\href {https://doi.org/10.1103/PhysRevB.92.161302} {\bibfield
			{journal} {\bibinfo  {journal} {Phys. Rev. B}\ }\textbf {\bibinfo {volume}
				{92}},\ \bibinfo {pages} {161302} (\bibinfo {year} {2015})}\BibitemShut
		{NoStop}%
		\bibitem [{\citenamefont {Weber}\ \emph {et~al.}(2018)\citenamefont {Weber},
			\citenamefont {Kettler}, \citenamefont {Vural}, \citenamefont {M{\"{u}}ller},
			\citenamefont {Maisch}, \citenamefont {Jetter}, \citenamefont {Portalupi},\
			and\ \citenamefont {Michler}}]{Weber2018c}%
		\BibitemOpen
		\bibfield  {author} {\bibinfo {author} {\bibfnamefont {J.~H.}\ \bibnamefont
				{Weber}}, \bibinfo {author} {\bibfnamefont {J.}~\bibnamefont {Kettler}},
			\bibinfo {author} {\bibfnamefont {H.}~\bibnamefont {Vural}}, \bibinfo
			{author} {\bibfnamefont {M.}~\bibnamefont {M{\"{u}}ller}}, \bibinfo {author}
			{\bibfnamefont {J.}~\bibnamefont {Maisch}}, \bibinfo {author} {\bibfnamefont
				{M.}~\bibnamefont {Jetter}}, \bibinfo {author} {\bibfnamefont {S.~L.}\
				\bibnamefont {Portalupi}},\ and\ \bibinfo {author} {\bibfnamefont
				{P.}~\bibnamefont {Michler}},\ }\bibfield  {title} {\bibinfo {title}
			{{Overcoming correlation fluctuations in two-photon interference experiments
					with differently bright and independently blinking remote quantum
					emitters}},\ }\href {https://doi.org/10.1103/PhysRevB.97.195414} {\bibfield
			{journal} {\bibinfo  {journal} {Phys. Rev. B}\ }\textbf {\bibinfo {volume}
				{97}},\ \bibinfo {pages} {195414} (\bibinfo {year} {2018})}\BibitemShut
		{NoStop}%
		\bibitem [{\citenamefont {Weber}\ \emph {et~al.}(2019)\citenamefont {Weber},
			\citenamefont {Kambs}, \citenamefont {Kettler}, \citenamefont {Kern},
			\citenamefont {Maisch}, \citenamefont {Vural}, \citenamefont {Jetter},
			\citenamefont {Portalupi}, \citenamefont {Becher},\ and\ \citenamefont
			{Michler}}]{Weber2019b}%
		\BibitemOpen
		\bibfield  {author} {\bibinfo {author} {\bibfnamefont {J.~H.}\ \bibnamefont
				{Weber}}, \bibinfo {author} {\bibfnamefont {B.}~\bibnamefont {Kambs}},
			\bibinfo {author} {\bibfnamefont {J.}~\bibnamefont {Kettler}}, \bibinfo
			{author} {\bibfnamefont {S.}~\bibnamefont {Kern}}, \bibinfo {author}
			{\bibfnamefont {J.}~\bibnamefont {Maisch}}, \bibinfo {author} {\bibfnamefont
				{H.}~\bibnamefont {Vural}}, \bibinfo {author} {\bibfnamefont
				{M.}~\bibnamefont {Jetter}}, \bibinfo {author} {\bibfnamefont {S.~L.}\
				\bibnamefont {Portalupi}}, \bibinfo {author} {\bibfnamefont {C.}~\bibnamefont
				{Becher}},\ and\ \bibinfo {author} {\bibfnamefont {P.}~\bibnamefont
				{Michler}},\ }\bibfield  {title} {\bibinfo {title} {{Two-photon interference
					in the telecom C-band after frequency conversion of photons from remote
					quantum emitters}},\ }\href {https://doi.org/10.1038/s41565-018-0279-8}
		{\bibfield  {journal} {\bibinfo  {journal} {Nat. Nanotechnol.}\ }\textbf
			{\bibinfo {volume} {14}},\ \bibinfo {pages} {23} (\bibinfo {year}
			{2019})}\BibitemShut {NoStop}%
		\bibitem [{\citenamefont {Zhai}\ \emph {et~al.}(2022)\citenamefont {Zhai},
			\citenamefont {Nguyen}, \citenamefont {Spinnler}, \citenamefont {Ritzmann},
			\citenamefont {L{\"{o}}bl}, \citenamefont {Wieck}, \citenamefont {Ludwig},
			\citenamefont {Javadi},\ and\ \citenamefont {Warburton}}]{Zhai2022a}%
		\BibitemOpen
		\bibfield  {author} {\bibinfo {author} {\bibfnamefont {L.}~\bibnamefont
				{Zhai}}, \bibinfo {author} {\bibfnamefont {G.~N.}\ \bibnamefont {Nguyen}},
			\bibinfo {author} {\bibfnamefont {C.}~\bibnamefont {Spinnler}}, \bibinfo
			{author} {\bibfnamefont {J.}~\bibnamefont {Ritzmann}}, \bibinfo {author}
			{\bibfnamefont {M.~C.}\ \bibnamefont {L{\"{o}}bl}}, \bibinfo {author}
			{\bibfnamefont {A.~D.}\ \bibnamefont {Wieck}}, \bibinfo {author}
			{\bibfnamefont {A.}~\bibnamefont {Ludwig}}, \bibinfo {author} {\bibfnamefont
				{A.}~\bibnamefont {Javadi}},\ and\ \bibinfo {author} {\bibfnamefont {R.~J.}\
				\bibnamefont {Warburton}},\ }\bibfield  {title} {\bibinfo {title} {{Quantum
					interference of identical photons from remote GaAs quantum dots}},\ }\href
		{https://doi.org/10.1038/s41565-022-01131-2} {\bibfield  {journal} {\bibinfo
				{journal} {Nat. Nanotechnol.}\ }\textbf {\bibinfo {volume} {17}},\ \bibinfo
			{pages} {829} (\bibinfo {year} {2022})}\BibitemShut {NoStop}%
		\bibitem [{\citenamefont {Akopian}\ \emph {et~al.}(2010)\citenamefont
			{Akopian}, \citenamefont {Perinetti}, \citenamefont {Wang}, \citenamefont
			{Rastelli}, \citenamefont {Schmidt},\ and\ \citenamefont
			{Zwiller}}]{Akopian2010}%
		\BibitemOpen
		\bibfield  {author} {\bibinfo {author} {\bibfnamefont {N.}~\bibnamefont
				{Akopian}}, \bibinfo {author} {\bibfnamefont {U.}~\bibnamefont {Perinetti}},
			\bibinfo {author} {\bibfnamefont {L.}~\bibnamefont {Wang}}, \bibinfo {author}
			{\bibfnamefont {A.}~\bibnamefont {Rastelli}}, \bibinfo {author}
			{\bibfnamefont {O.~G.}\ \bibnamefont {Schmidt}},\ and\ \bibinfo {author}
			{\bibfnamefont {V.}~\bibnamefont {Zwiller}},\ }\bibfield  {title} {\bibinfo
			{title} {{Tuning single GaAs quantum dots in resonance with a rubidium
					vapor}},\ }\href {https://doi.org/10.1063/1.3478232} {\bibfield  {journal}
			{\bibinfo  {journal} {Appl. Phys. Lett.}\ }\textbf {\bibinfo {volume} {97}},\
			\bibinfo {pages} {082103} (\bibinfo {year} {2010})}\BibitemShut {NoStop}%
		\bibitem [{\citenamefont {Trotta}\ \emph {et~al.}(2012)\citenamefont {Trotta},
			\citenamefont {Zallo}, \citenamefont {Ortix}, \citenamefont {Atkinson},
			\citenamefont {Plumhof}, \citenamefont {van~den Brink}, \citenamefont
			{Rastelli},\ and\ \citenamefont {Schmidt}}]{Trotta2012}%
		\BibitemOpen
		\bibfield  {author} {\bibinfo {author} {\bibfnamefont {R.}~\bibnamefont
				{Trotta}}, \bibinfo {author} {\bibfnamefont {E.}~\bibnamefont {Zallo}},
			\bibinfo {author} {\bibfnamefont {C.}~\bibnamefont {Ortix}}, \bibinfo
			{author} {\bibfnamefont {P.}~\bibnamefont {Atkinson}}, \bibinfo {author}
			{\bibfnamefont {J.~D.}\ \bibnamefont {Plumhof}}, \bibinfo {author}
			{\bibfnamefont {J.}~\bibnamefont {van~den Brink}}, \bibinfo {author}
			{\bibfnamefont {A.}~\bibnamefont {Rastelli}},\ and\ \bibinfo {author}
			{\bibfnamefont {O.~G.}\ \bibnamefont {Schmidt}},\ }\bibfield  {title}
		{\bibinfo {title} {{Universal Recovery of the Energy-Level Degeneracy of
					Bright Excitons in InGaAs Quantum Dots without a Structure Symmetry}},\
		}\href {https://doi.org/10.1103/PhysRevLett.109.147401} {\bibfield  {journal}
			{\bibinfo  {journal} {Phys. Rev. Lett.}\ }\textbf {\bibinfo {volume} {109}},\
			\bibinfo {pages} {147401} (\bibinfo {year} {2012})}\BibitemShut {NoStop}%
		\bibitem [{\citenamefont {Mart{\'{i}}n-S{\'{a}}nchez}\ \emph
			{et~al.}(2018)\citenamefont {Mart{\'{i}}n-S{\'{a}}nchez}, \citenamefont
			{Trotta}, \citenamefont {Mariscal}, \citenamefont {Serna}, \citenamefont
			{Piredda}, \citenamefont {Stroj}, \citenamefont {Edlinger}, \citenamefont
			{Schimpf}, \citenamefont {Aberl}, \citenamefont {Lettner}, \citenamefont
			{Wildmann}, \citenamefont {Huang}, \citenamefont {Yuan}, \citenamefont
			{Ziss}, \citenamefont {Stangl},\ and\ \citenamefont
			{Rastelli}}]{Martin-Sanchez2018}%
		\BibitemOpen
		\bibfield  {author} {\bibinfo {author} {\bibfnamefont {J.}~\bibnamefont
				{Mart{\'{i}}n-S{\'{a}}nchez}}, \bibinfo {author} {\bibfnamefont
				{R.}~\bibnamefont {Trotta}}, \bibinfo {author} {\bibfnamefont
				{A.}~\bibnamefont {Mariscal}}, \bibinfo {author} {\bibfnamefont
				{R.}~\bibnamefont {Serna}}, \bibinfo {author} {\bibfnamefont
				{G.}~\bibnamefont {Piredda}}, \bibinfo {author} {\bibfnamefont
				{S.}~\bibnamefont {Stroj}}, \bibinfo {author} {\bibfnamefont
				{J.}~\bibnamefont {Edlinger}}, \bibinfo {author} {\bibfnamefont
				{C.}~\bibnamefont {Schimpf}}, \bibinfo {author} {\bibfnamefont
				{J.}~\bibnamefont {Aberl}}, \bibinfo {author} {\bibfnamefont
				{T.}~\bibnamefont {Lettner}}, \bibinfo {author} {\bibfnamefont
				{J.}~\bibnamefont {Wildmann}}, \bibinfo {author} {\bibfnamefont
				{H.}~\bibnamefont {Huang}}, \bibinfo {author} {\bibfnamefont
				{X.}~\bibnamefont {Yuan}}, \bibinfo {author} {\bibfnamefont {D.}~\bibnamefont
				{Ziss}}, \bibinfo {author} {\bibfnamefont {J.}~\bibnamefont {Stangl}},\ and\
			\bibinfo {author} {\bibfnamefont {A.}~\bibnamefont {Rastelli}},\ }\bibfield
		{title} {\bibinfo {title} {{Strain-tuning of the optical properties of
					semiconductor nanomaterials by integration onto piezoelectric actuators}},\
		}\href {https://doi.org/10.1088/1361-6641/aa9b53} {\bibfield  {journal}
			{\bibinfo  {journal} {Semicond. Sci. Technol.}\ }\textbf {\bibinfo {volume}
				{33}},\ \bibinfo {pages} {013001} (\bibinfo {year} {2018})}\BibitemShut
		{NoStop}%
		\bibitem [{\citenamefont {Wang}\ \emph {et~al.}(2016)\citenamefont {Wang},
			\citenamefont {Duan}, \citenamefont {Li}, \citenamefont {Chen}, \citenamefont
			{Li}, \citenamefont {He}, \citenamefont {Chen}, \citenamefont {He},
			\citenamefont {Ding}, \citenamefont {Peng}, \citenamefont {Schneider},
			\citenamefont {Kamp}, \citenamefont {H{\"{o}}fling}, \citenamefont {Lu},\
			and\ \citenamefont {Pan}}]{Wang2016a}%
		\BibitemOpen
		\bibfield  {author} {\bibinfo {author} {\bibfnamefont {H.}~\bibnamefont
				{Wang}}, \bibinfo {author} {\bibfnamefont {Z.-C.}\ \bibnamefont {Duan}},
			\bibinfo {author} {\bibfnamefont {Y.-H.}\ \bibnamefont {Li}}, \bibinfo
			{author} {\bibfnamefont {S.}~\bibnamefont {Chen}}, \bibinfo {author}
			{\bibfnamefont {J.-P.}\ \bibnamefont {Li}}, \bibinfo {author} {\bibfnamefont
				{Y.-M.}\ \bibnamefont {He}}, \bibinfo {author} {\bibfnamefont {M.-C.}\
				\bibnamefont {Chen}}, \bibinfo {author} {\bibfnamefont {Y.}~\bibnamefont
				{He}}, \bibinfo {author} {\bibfnamefont {X.}~\bibnamefont {Ding}}, \bibinfo
			{author} {\bibfnamefont {C.-Z.}\ \bibnamefont {Peng}}, \bibinfo {author}
			{\bibfnamefont {C.}~\bibnamefont {Schneider}}, \bibinfo {author}
			{\bibfnamefont {M.}~\bibnamefont {Kamp}}, \bibinfo {author} {\bibfnamefont
				{S.}~\bibnamefont {H{\"{o}}fling}}, \bibinfo {author} {\bibfnamefont {C.-Y.}\
				\bibnamefont {Lu}},\ and\ \bibinfo {author} {\bibfnamefont {J.-W.}\
				\bibnamefont {Pan}},\ }\bibfield  {title} {\bibinfo {title}
			{{Near-Transform-Limited Single Photons from an Efficient Solid-State Quantum
					Emitter}},\ }\href {https://doi.org/10.1103/PhysRevLett.116.213601}
		{\bibfield  {journal} {\bibinfo  {journal} {Phys. Rev. Lett.}\ }\textbf
			{\bibinfo {volume} {116}},\ \bibinfo {pages} {213601} (\bibinfo {year}
			{2016})}\BibitemShut {NoStop}%
		\bibitem [{\citenamefont {Wang}\ \emph {et~al.}(2019)\citenamefont {Wang},
			\citenamefont {Hu}, \citenamefont {Chung}, \citenamefont {Qin}, \citenamefont
			{Yang}, \citenamefont {Li}, \citenamefont {Liu}, \citenamefont {Zhong},
			\citenamefont {He}, \citenamefont {Ding}, \citenamefont {Deng}, \citenamefont
			{Dai}, \citenamefont {Huo}, \citenamefont {H{\"{o}}fling}, \citenamefont
			{Lu},\ and\ \citenamefont {Pan}}]{Wang2019h}%
		\BibitemOpen
		\bibfield  {author} {\bibinfo {author} {\bibfnamefont {H.}~\bibnamefont
				{Wang}}, \bibinfo {author} {\bibfnamefont {H.}~\bibnamefont {Hu}}, \bibinfo
			{author} {\bibfnamefont {T.-H.}\ \bibnamefont {Chung}}, \bibinfo {author}
			{\bibfnamefont {J.}~\bibnamefont {Qin}}, \bibinfo {author} {\bibfnamefont
				{X.}~\bibnamefont {Yang}}, \bibinfo {author} {\bibfnamefont {J.-P.}\
				\bibnamefont {Li}}, \bibinfo {author} {\bibfnamefont {R.-Z.}\ \bibnamefont
				{Liu}}, \bibinfo {author} {\bibfnamefont {H.-S.}\ \bibnamefont {Zhong}},
			\bibinfo {author} {\bibfnamefont {Y.-M.}\ \bibnamefont {He}}, \bibinfo
			{author} {\bibfnamefont {X.}~\bibnamefont {Ding}}, \bibinfo {author}
			{\bibfnamefont {Y.-H.}\ \bibnamefont {Deng}}, \bibinfo {author}
			{\bibfnamefont {Q.}~\bibnamefont {Dai}}, \bibinfo {author} {\bibfnamefont
				{Y.-H.}\ \bibnamefont {Huo}}, \bibinfo {author} {\bibfnamefont
				{S.}~\bibnamefont {H{\"{o}}fling}}, \bibinfo {author} {\bibfnamefont {C.-Y.}\
				\bibnamefont {Lu}},\ and\ \bibinfo {author} {\bibfnamefont {J.-W.}\
				\bibnamefont {Pan}},\ }\bibfield  {title} {\bibinfo {title} {{On-Demand
					Semiconductor Source of Entangled Photons Which Simultaneously Has High
					Fidelity, Efficiency, and Indistinguishability}},\ }\href
		{https://doi.org/10.1103/PhysRevLett.122.113602} {\bibfield  {journal}
			{\bibinfo  {journal} {Phys. Rev. Lett.}\ }\textbf {\bibinfo {volume} {122}},\
			\bibinfo {pages} {113602} (\bibinfo {year} {2019})}\BibitemShut {NoStop}%
		\bibitem [{\citenamefont {Liu}\ \emph {et~al.}(2019)\citenamefont {Liu},
			\citenamefont {Su}, \citenamefont {Wei}, \citenamefont {Yao}, \citenamefont
			{da~Silva}, \citenamefont {Yu}, \citenamefont {Iles-Smith}, \citenamefont
			{Srinivasan}, \citenamefont {Rastelli}, \citenamefont {Li},\ and\
			\citenamefont {Wang}}]{Liu2019b}%
		\BibitemOpen
		\bibfield  {author} {\bibinfo {author} {\bibfnamefont {J.}~\bibnamefont
				{Liu}}, \bibinfo {author} {\bibfnamefont {R.}~\bibnamefont {Su}}, \bibinfo
			{author} {\bibfnamefont {Y.}~\bibnamefont {Wei}}, \bibinfo {author}
			{\bibfnamefont {B.}~\bibnamefont {Yao}}, \bibinfo {author} {\bibfnamefont
				{S.~F.~C.}\ \bibnamefont {da~Silva}}, \bibinfo {author} {\bibfnamefont
				{Y.}~\bibnamefont {Yu}}, \bibinfo {author} {\bibfnamefont {J.}~\bibnamefont
				{Iles-Smith}}, \bibinfo {author} {\bibfnamefont {K.}~\bibnamefont
				{Srinivasan}}, \bibinfo {author} {\bibfnamefont {A.}~\bibnamefont
				{Rastelli}}, \bibinfo {author} {\bibfnamefont {J.}~\bibnamefont {Li}},\ and\
			\bibinfo {author} {\bibfnamefont {X.}~\bibnamefont {Wang}},\ }\bibfield
		{title} {\bibinfo {title} {{A solid-state source of strongly entangled photon
					pairs with high brightness and indistinguishability}},\ }\href
		{https://doi.org/10.1038/s41565-019-0435-9} {\bibfield  {journal} {\bibinfo
				{journal} {Nat. Nanotechnol.}\ }\textbf {\bibinfo {volume} {14}},\ \bibinfo
			{pages} {586} (\bibinfo {year} {2019})}\BibitemShut {NoStop}%
		\bibitem [{\citenamefont {Somaschi}\ \emph {et~al.}(2016)\citenamefont
			{Somaschi}, \citenamefont {Giesz}, \citenamefont {{De Santis}}, \citenamefont
			{Loredo}, \citenamefont {Almeida}, \citenamefont {Hornecker}, \citenamefont
			{Portalupi}, \citenamefont {Grange}, \citenamefont {Ant{\'{o}}n},
			\citenamefont {Demory}, \citenamefont {G{\'{o}}mez}, \citenamefont {Sagnes},
			\citenamefont {Lanzillotti-Kimura}, \citenamefont {Lema{\'{i}}tre},
			\citenamefont {Auffeves}, \citenamefont {White}, \citenamefont {Lanco},\ and\
			\citenamefont {Senellart}}]{Somaschi2016a}%
		\BibitemOpen
		\bibfield  {author} {\bibinfo {author} {\bibfnamefont {N.}~\bibnamefont
				{Somaschi}}, \bibinfo {author} {\bibfnamefont {V.}~\bibnamefont {Giesz}},
			\bibinfo {author} {\bibfnamefont {L.}~\bibnamefont {{De Santis}}}, \bibinfo
			{author} {\bibfnamefont {J.~C.}\ \bibnamefont {Loredo}}, \bibinfo {author}
			{\bibfnamefont {M.~P.}\ \bibnamefont {Almeida}}, \bibinfo {author}
			{\bibfnamefont {G.}~\bibnamefont {Hornecker}}, \bibinfo {author}
			{\bibfnamefont {S.~L.}\ \bibnamefont {Portalupi}}, \bibinfo {author}
			{\bibfnamefont {T.}~\bibnamefont {Grange}}, \bibinfo {author} {\bibfnamefont
				{C.}~\bibnamefont {Ant{\'{o}}n}}, \bibinfo {author} {\bibfnamefont
				{J.}~\bibnamefont {Demory}}, \bibinfo {author} {\bibfnamefont
				{C.}~\bibnamefont {G{\'{o}}mez}}, \bibinfo {author} {\bibfnamefont
				{I.}~\bibnamefont {Sagnes}}, \bibinfo {author} {\bibfnamefont {N.~D.}\
				\bibnamefont {Lanzillotti-Kimura}}, \bibinfo {author} {\bibfnamefont
				{A.}~\bibnamefont {Lema{\'{i}}tre}}, \bibinfo {author} {\bibfnamefont
				{A.}~\bibnamefont {Auffeves}}, \bibinfo {author} {\bibfnamefont {A.~G.}\
				\bibnamefont {White}}, \bibinfo {author} {\bibfnamefont {L.}~\bibnamefont
				{Lanco}},\ and\ \bibinfo {author} {\bibfnamefont {P.}~\bibnamefont
				{Senellart}},\ }\bibfield  {title} {\bibinfo {title} {{Near-optimal
					single-photon sources in the solid state}},\ }\href
		{https://doi.org/10.1038/nphoton.2016.23} {\bibfield  {journal} {\bibinfo
				{journal} {Nat. Photonics}\ }\textbf {\bibinfo {volume} {10}},\ \bibinfo
			{pages} {340} (\bibinfo {year} {2016})}\BibitemShut {NoStop}%
		\bibitem [{\citenamefont {Kuhlmann}\ \emph {et~al.}(2015)\citenamefont
			{Kuhlmann}, \citenamefont {Prechtel}, \citenamefont {Houel}, \citenamefont
			{Ludwig}, \citenamefont {Reuter}, \citenamefont {Wieck},\ and\ \citenamefont
			{Warburton}}]{Kuhlmann2015a}%
		\BibitemOpen
		\bibfield  {author} {\bibinfo {author} {\bibfnamefont {A.~V.}\ \bibnamefont
				{Kuhlmann}}, \bibinfo {author} {\bibfnamefont {J.~H.}\ \bibnamefont
				{Prechtel}}, \bibinfo {author} {\bibfnamefont {J.}~\bibnamefont {Houel}},
			\bibinfo {author} {\bibfnamefont {A.}~\bibnamefont {Ludwig}}, \bibinfo
			{author} {\bibfnamefont {D.}~\bibnamefont {Reuter}}, \bibinfo {author}
			{\bibfnamefont {A.~D.}\ \bibnamefont {Wieck}},\ and\ \bibinfo {author}
			{\bibfnamefont {R.~J.}\ \bibnamefont {Warburton}},\ }\bibfield  {title}
		{\bibinfo {title} {{Transform-limited single photons from a single quantum
					dot}},\ }\href {https://doi.org/10.1038/ncomms9204} {\bibfield  {journal}
			{\bibinfo  {journal} {Nat. Commun.}\ }\textbf {\bibinfo {volume} {6}},\
			\bibinfo {pages} {8204} (\bibinfo {year} {2015})}\BibitemShut {NoStop}%
		\bibitem [{\citenamefont {Prechtel}\ \emph {et~al.}(2016)\citenamefont
			{Prechtel}, \citenamefont {Kuhlmann}, \citenamefont {Houel}, \citenamefont
			{Ludwig}, \citenamefont {Valentin}, \citenamefont {Wieck},\ and\
			\citenamefont {Warburton}}]{Prechtel2016a}%
		\BibitemOpen
		\bibfield  {author} {\bibinfo {author} {\bibfnamefont {J.~H.}\ \bibnamefont
				{Prechtel}}, \bibinfo {author} {\bibfnamefont {A.~V.}\ \bibnamefont
				{Kuhlmann}}, \bibinfo {author} {\bibfnamefont {J.}~\bibnamefont {Houel}},
			\bibinfo {author} {\bibfnamefont {A.}~\bibnamefont {Ludwig}}, \bibinfo
			{author} {\bibfnamefont {S.~R.}\ \bibnamefont {Valentin}}, \bibinfo {author}
			{\bibfnamefont {A.~D.}\ \bibnamefont {Wieck}},\ and\ \bibinfo {author}
			{\bibfnamefont {R.~J.}\ \bibnamefont {Warburton}},\ }\bibfield  {title}
		{\bibinfo {title} {{Decoupling a hole spin qubit from the nuclear spins}},\
		}\href {https://doi.org/10.1038/nmat4704} {\bibfield  {journal} {\bibinfo
				{journal} {Nat. Mater.}\ }\textbf {\bibinfo {volume} {15}},\ \bibinfo {pages}
			{981} (\bibinfo {year} {2016})}\BibitemShut {NoStop}%
		\bibitem [{\citenamefont {Tomm}\ \emph {et~al.}(2021)\citenamefont {Tomm},
			\citenamefont {Javadi}, \citenamefont {Antoniadis}, \citenamefont {Najer},
			\citenamefont {L{\"{o}}bl}, \citenamefont {Korsch}, \citenamefont {Schott},
			\citenamefont {Valentin}, \citenamefont {Wieck}, \citenamefont {Ludwig},\
			and\ \citenamefont {Warburton}}]{Tomm2021a}%
		\BibitemOpen
		\bibfield  {author} {\bibinfo {author} {\bibfnamefont {N.}~\bibnamefont
				{Tomm}}, \bibinfo {author} {\bibfnamefont {A.}~\bibnamefont {Javadi}},
			\bibinfo {author} {\bibfnamefont {N.~O.}\ \bibnamefont {Antoniadis}},
			\bibinfo {author} {\bibfnamefont {D.}~\bibnamefont {Najer}}, \bibinfo
			{author} {\bibfnamefont {M.~C.}\ \bibnamefont {L{\"{o}}bl}}, \bibinfo
			{author} {\bibfnamefont {A.~R.}\ \bibnamefont {Korsch}}, \bibinfo {author}
			{\bibfnamefont {R.}~\bibnamefont {Schott}}, \bibinfo {author} {\bibfnamefont
				{S.~R.}\ \bibnamefont {Valentin}}, \bibinfo {author} {\bibfnamefont {A.~D.}\
				\bibnamefont {Wieck}}, \bibinfo {author} {\bibfnamefont {A.}~\bibnamefont
				{Ludwig}},\ and\ \bibinfo {author} {\bibfnamefont {R.~J.}\ \bibnamefont
				{Warburton}},\ }\bibfield  {title} {\bibinfo {title} {{A bright and fast
					source of coherent single photons}},\ }\href
		{https://doi.org/10.1038/s41565-020-00831-x} {\bibfield  {journal} {\bibinfo
				{journal} {Nat. Nanotechnol.}\ }\textbf {\bibinfo {volume} {16}},\ \bibinfo
			{pages} {399} (\bibinfo {year} {2021})}\BibitemShut {NoStop}%
		\bibitem [{\citenamefont {Reuter}\ \emph {et~al.}(1999)\citenamefont {Reuter},
			\citenamefont {Wieck},\ and\ \citenamefont {Fischer}}]{Reuter1999}%
		\BibitemOpen
		\bibfield  {author} {\bibinfo {author} {\bibfnamefont {D.}~\bibnamefont
				{Reuter}}, \bibinfo {author} {\bibfnamefont {A.~D.}\ \bibnamefont {Wieck}},\
			and\ \bibinfo {author} {\bibfnamefont {A.}~\bibnamefont {Fischer}},\
		}\bibfield  {title} {\bibinfo {title} {{A compact electron beam evaporator
					for carbon doping in solid source molecular beam epitaxy}},\ }\href
		{https://doi.org/10.1063/1.1149933} {\bibfield  {journal} {\bibinfo
				{journal} {Rev. Sci. Instrum.}\ }\textbf {\bibinfo {volume} {70}},\ \bibinfo
			{pages} {3435} (\bibinfo {year} {1999})}\BibitemShut {NoStop}%
		\bibitem [{\citenamefont {Ludwig}\ \emph {et~al.}(2017)\citenamefont {Ludwig},
			\citenamefont {Prechtel}, \citenamefont {Kuhlmann}, \citenamefont {Houel},
			\citenamefont {Valentin}, \citenamefont {Warburton},\ and\ \citenamefont
			{Wieck}}]{Ludwig2017}%
		\BibitemOpen
		\bibfield  {author} {\bibinfo {author} {\bibfnamefont {A.}~\bibnamefont
				{Ludwig}}, \bibinfo {author} {\bibfnamefont {J.~H.}\ \bibnamefont
				{Prechtel}}, \bibinfo {author} {\bibfnamefont {A.~V.}\ \bibnamefont
				{Kuhlmann}}, \bibinfo {author} {\bibfnamefont {J.}~\bibnamefont {Houel}},
			\bibinfo {author} {\bibfnamefont {S.~R.}\ \bibnamefont {Valentin}}, \bibinfo
			{author} {\bibfnamefont {R.~J.}\ \bibnamefont {Warburton}},\ and\ \bibinfo
			{author} {\bibfnamefont {A.~D.}\ \bibnamefont {Wieck}},\ }\bibfield  {title}
		{\bibinfo {title} {{Ultra-low charge and spin noise in self-assembled quantum
					dots}},\ }\href {https://doi.org/10.1016/j.jcrysgro.2017.05.008} {\bibfield
			{journal} {\bibinfo  {journal} {J. Cryst. Growth}\ }\textbf {\bibinfo
				{volume} {477}},\ \bibinfo {pages} {193} (\bibinfo {year}
			{2017})}\BibitemShut {NoStop}%
		\bibitem [{\citenamefont {Pedersen}\ \emph {et~al.}(2020)\citenamefont
			{Pedersen}, \citenamefont {Wang}, \citenamefont {Olesen}, \citenamefont
			{Scholz}, \citenamefont {Wieck}, \citenamefont {Ludwig}, \citenamefont
			{L{\"{o}}bl}, \citenamefont {Warburton}, \citenamefont {Midolo},
			\citenamefont {Uppu},\ and\ \citenamefont {Lodahl}}]{Pedersen2020a}%
		\BibitemOpen
		\bibfield  {author} {\bibinfo {author} {\bibfnamefont {F.~T.}\ \bibnamefont
				{Pedersen}}, \bibinfo {author} {\bibfnamefont {Y.}~\bibnamefont {Wang}},
			\bibinfo {author} {\bibfnamefont {C.~T.}\ \bibnamefont {Olesen}}, \bibinfo
			{author} {\bibfnamefont {S.}~\bibnamefont {Scholz}}, \bibinfo {author}
			{\bibfnamefont {A.~D.}\ \bibnamefont {Wieck}}, \bibinfo {author}
			{\bibfnamefont {A.}~\bibnamefont {Ludwig}}, \bibinfo {author} {\bibfnamefont
				{M.~C.}\ \bibnamefont {L{\"{o}}bl}}, \bibinfo {author} {\bibfnamefont
				{R.~J.}\ \bibnamefont {Warburton}}, \bibinfo {author} {\bibfnamefont
				{L.}~\bibnamefont {Midolo}}, \bibinfo {author} {\bibfnamefont
				{R.}~\bibnamefont {Uppu}},\ and\ \bibinfo {author} {\bibfnamefont
				{P.}~\bibnamefont {Lodahl}},\ }\bibfield  {title} {\bibinfo {title} {{Near
					Transform-Limited Quantum Dot Linewidths in a Broadband Photonic Crystal
					Waveguide}},\ }\href {https://doi.org/10.1021/acsphotonics.0c00758}
		{\bibfield  {journal} {\bibinfo  {journal} {ACS Photonics}\ }\textbf
			{\bibinfo {volume} {7}},\ \bibinfo {pages} {2343} (\bibinfo {year}
			{2020})}\BibitemShut {NoStop}%
		\bibitem [{\citenamefont {Bauch}\ \emph {et~al.}(2021)\citenamefont {Bauch},
			\citenamefont {Heinze}, \citenamefont {F{\"{o}}rstner}, \citenamefont
			{J{\"{o}}ns},\ and\ \citenamefont {Schumacher}}]{Bauch2021b}%
		\BibitemOpen
		\bibfield  {author} {\bibinfo {author} {\bibfnamefont {D.}~\bibnamefont
				{Bauch}}, \bibinfo {author} {\bibfnamefont {D.}~\bibnamefont {Heinze}},
			\bibinfo {author} {\bibfnamefont {J.}~\bibnamefont {F{\"{o}}rstner}},
			\bibinfo {author} {\bibfnamefont {K.~D.}\ \bibnamefont {J{\"{o}}ns}},\ and\
			\bibinfo {author} {\bibfnamefont {S.}~\bibnamefont {Schumacher}},\ }\bibfield
		{title} {\bibinfo {title} {{Ultrafast electric control of cavity mediated
					single-photon and photon-pair generation with semiconductor quantum dots}},\
		}\href {https://doi.org/10.1103/PhysRevB.104.085308} {\bibfield  {journal}
			{\bibinfo  {journal} {Phys. Rev. B}\ }\textbf {\bibinfo {volume} {104}},\
			\bibinfo {pages} {085308} (\bibinfo {year} {2021})}\BibitemShut {NoStop}%
		\bibitem [{\citenamefont {McCray}(2007)}]{Mccray2007}%
		\BibitemOpen
		\bibfield  {author} {\bibinfo {author} {\bibfnamefont {W.~P.}\ \bibnamefont
				{McCray}},\ }\bibfield  {title} {\bibinfo {title} {{MBE deserves a place in
					the history books}},\ }\href {https://doi.org/10.1038/nnano.2007.121}
		{\bibfield  {journal} {\bibinfo  {journal} {Nat. Nanotechnol.}\ }\textbf
			{\bibinfo {volume} {2}},\ \bibinfo {pages} {259} (\bibinfo {year}
			{2007})}\BibitemShut {NoStop}%
		\bibitem [{\citenamefont {Najer}\ \emph {et~al.}(2019)\citenamefont {Najer},
			\citenamefont {S{\"{o}}llner}, \citenamefont {Sekatski}, \citenamefont
			{Dolique}, \citenamefont {L{\"{o}}bl}, \citenamefont {Riedel}, \citenamefont
			{Schott}, \citenamefont {Starosielec}, \citenamefont {Valentin},
			\citenamefont {Wieck}, \citenamefont {Sangouard}, \citenamefont {Ludwig},\
			and\ \citenamefont {Warburton}}]{Najer2019b}%
		\BibitemOpen
		\bibfield  {author} {\bibinfo {author} {\bibfnamefont {D.}~\bibnamefont
				{Najer}}, \bibinfo {author} {\bibfnamefont {I.}~\bibnamefont
				{S{\"{o}}llner}}, \bibinfo {author} {\bibfnamefont {P.}~\bibnamefont
				{Sekatski}}, \bibinfo {author} {\bibfnamefont {V.}~\bibnamefont {Dolique}},
			\bibinfo {author} {\bibfnamefont {M.~C.}\ \bibnamefont {L{\"{o}}bl}},
			\bibinfo {author} {\bibfnamefont {D.}~\bibnamefont {Riedel}}, \bibinfo
			{author} {\bibfnamefont {R.}~\bibnamefont {Schott}}, \bibinfo {author}
			{\bibfnamefont {S.}~\bibnamefont {Starosielec}}, \bibinfo {author}
			{\bibfnamefont {S.~R.}\ \bibnamefont {Valentin}}, \bibinfo {author}
			{\bibfnamefont {A.~D.}\ \bibnamefont {Wieck}}, \bibinfo {author}
			{\bibfnamefont {N.}~\bibnamefont {Sangouard}}, \bibinfo {author}
			{\bibfnamefont {A.}~\bibnamefont {Ludwig}},\ and\ \bibinfo {author}
			{\bibfnamefont {R.~J.}\ \bibnamefont {Warburton}},\ }\bibfield  {title}
		{\bibinfo {title} {{A gated quantum dot strongly coupled to an optical
					microcavity}},\ }\href {https://doi.org/10.1038/s41586-019-1709-y} {\bibfield
			{journal} {\bibinfo  {journal} {Nature}\ }\textbf {\bibinfo {volume}
				{575}},\ \bibinfo {pages} {622} (\bibinfo {year} {2019})}\BibitemShut
		{NoStop}%
		\bibitem [{\citenamefont {Kosarev}\ \emph {et~al.}(2022)\citenamefont
			{Kosarev}, \citenamefont {Trifonov}, \citenamefont {Yugova}, \citenamefont
			{Yanibekov}, \citenamefont {Poltavtsev}, \citenamefont {Kamenskii},
			\citenamefont {Scholz}, \citenamefont {Sgroi}, \citenamefont {Ludwig},
			\citenamefont {Wieck}, \citenamefont {Yakovlev}, \citenamefont {Bayer},\ and\
			\citenamefont {Akimov}}]{Kosarev2022}%
		\BibitemOpen
		\bibfield  {author} {\bibinfo {author} {\bibfnamefont {A.~N.}\ \bibnamefont
				{Kosarev}}, \bibinfo {author} {\bibfnamefont {A.~V.}\ \bibnamefont
				{Trifonov}}, \bibinfo {author} {\bibfnamefont {I.~A.}\ \bibnamefont
				{Yugova}}, \bibinfo {author} {\bibfnamefont {I.~I.}\ \bibnamefont
				{Yanibekov}}, \bibinfo {author} {\bibfnamefont {S.~V.}\ \bibnamefont
				{Poltavtsev}}, \bibinfo {author} {\bibfnamefont {A.~N.}\ \bibnamefont
				{Kamenskii}}, \bibinfo {author} {\bibfnamefont {S.~E.}\ \bibnamefont
				{Scholz}}, \bibinfo {author} {\bibfnamefont {C.~A.}\ \bibnamefont {Sgroi}},
			\bibinfo {author} {\bibfnamefont {A.}~\bibnamefont {Ludwig}}, \bibinfo
			{author} {\bibfnamefont {A.~D.}\ \bibnamefont {Wieck}}, \bibinfo {author}
			{\bibfnamefont {D.~R.}\ \bibnamefont {Yakovlev}}, \bibinfo {author}
			{\bibfnamefont {M.}~\bibnamefont {Bayer}},\ and\ \bibinfo {author}
			{\bibfnamefont {I.~A.}\ \bibnamefont {Akimov}},\ }\bibfield  {title}
		{\bibinfo {title} {{Extending the time of coherent optical response in
					ensemble of singly-charged InGaAs quantum dots}},\ }\href
		{https://doi.org/10.1038/s42005-022-00922-2} {\bibfield  {journal} {\bibinfo
				{journal} {Commun. Phys.}\ }\textbf {\bibinfo {volume} {5}},\ \bibinfo
			{pages} {144} (\bibinfo {year} {2022})}\BibitemShut {NoStop}%
		\bibitem [{\citenamefont {Brokmann}\ \emph {et~al.}(2006)\citenamefont
			{Brokmann}, \citenamefont {Bawendi}, \citenamefont {Coolen},\ and\
			\citenamefont {Hermier}}]{Brokmann2006b}%
		\BibitemOpen
		\bibfield  {author} {\bibinfo {author} {\bibfnamefont {X.}~\bibnamefont
				{Brokmann}}, \bibinfo {author} {\bibfnamefont {M.}~\bibnamefont {Bawendi}},
			\bibinfo {author} {\bibfnamefont {L.}~\bibnamefont {Coolen}},\ and\ \bibinfo
			{author} {\bibfnamefont {J.-P.}\ \bibnamefont {Hermier}},\ }\bibfield
		{title} {\bibinfo {title} {{Photon-correlation Fourier spectroscopy}},\
		}\href {https://doi.org/10.1364/OE.14.006333} {\bibfield  {journal} {\bibinfo
				{journal} {Opt. Express}\ }\textbf {\bibinfo {volume} {14}},\ \bibinfo
			{pages} {6333} (\bibinfo {year} {2006})}\BibitemShut {NoStop}%
		\bibitem [{\citenamefont {Schimpf}\ \emph {et~al.}(2019)\citenamefont
			{Schimpf}, \citenamefont {Reindl}, \citenamefont {Klenovsk{\'{y}}},
			\citenamefont {Fromherz}, \citenamefont {{Covre Da Silva}}, \citenamefont
			{Hofer}, \citenamefont {Schneider}, \citenamefont {H{\"{o}}fling},
			\citenamefont {Trotta},\ and\ \citenamefont {Rastelli}}]{Schimpf2019b}%
		\BibitemOpen
		\bibfield  {author} {\bibinfo {author} {\bibfnamefont {C.}~\bibnamefont
				{Schimpf}}, \bibinfo {author} {\bibfnamefont {M.}~\bibnamefont {Reindl}},
			\bibinfo {author} {\bibfnamefont {P.}~\bibnamefont {Klenovsk{\'{y}}}},
			\bibinfo {author} {\bibfnamefont {T.}~\bibnamefont {Fromherz}}, \bibinfo
			{author} {\bibfnamefont {S.~F.}\ \bibnamefont {{Covre Da Silva}}}, \bibinfo
			{author} {\bibfnamefont {J.}~\bibnamefont {Hofer}}, \bibinfo {author}
			{\bibfnamefont {C.}~\bibnamefont {Schneider}}, \bibinfo {author}
			{\bibfnamefont {S.}~\bibnamefont {H{\"{o}}fling}}, \bibinfo {author}
			{\bibfnamefont {R.}~\bibnamefont {Trotta}},\ and\ \bibinfo {author}
			{\bibfnamefont {A.}~\bibnamefont {Rastelli}},\ }\bibfield  {title} {\bibinfo
			{title} {{Resolving the temporal evolution of line broadening in single
					quantum emitters}},\ }\href {https://doi.org/10.1364/OE.27.035290} {\bibfield
			{journal} {\bibinfo  {journal} {Opt. Express}\ }\textbf {\bibinfo {volume}
				{27}},\ \bibinfo {pages} {35290} (\bibinfo {year} {2019})}\BibitemShut
		{NoStop}%
		\bibitem [{\citenamefont {L{\"{o}}bl}\ \emph {et~al.}(2017)\citenamefont
			{L{\"{o}}bl}, \citenamefont {S{\"{o}}llner}, \citenamefont {Javadi},
			\citenamefont {Pregnolato}, \citenamefont {Schott}, \citenamefont {Midolo},
			\citenamefont {Kuhlmann}, \citenamefont {Stobbe}, \citenamefont {Wieck},
			\citenamefont {Lodahl}, \citenamefont {Ludwig},\ and\ \citenamefont
			{Warburton}}]{Lobl2017a}%
		\BibitemOpen
		\bibfield  {author} {\bibinfo {author} {\bibfnamefont {M.~C.}\ \bibnamefont
				{L{\"{o}}bl}}, \bibinfo {author} {\bibfnamefont {I.}~\bibnamefont
				{S{\"{o}}llner}}, \bibinfo {author} {\bibfnamefont {A.}~\bibnamefont
				{Javadi}}, \bibinfo {author} {\bibfnamefont {T.}~\bibnamefont {Pregnolato}},
			\bibinfo {author} {\bibfnamefont {R.}~\bibnamefont {Schott}}, \bibinfo
			{author} {\bibfnamefont {L.}~\bibnamefont {Midolo}}, \bibinfo {author}
			{\bibfnamefont {A.~V.}\ \bibnamefont {Kuhlmann}}, \bibinfo {author}
			{\bibfnamefont {S.}~\bibnamefont {Stobbe}}, \bibinfo {author} {\bibfnamefont
				{A.~D.}\ \bibnamefont {Wieck}}, \bibinfo {author} {\bibfnamefont
				{P.}~\bibnamefont {Lodahl}}, \bibinfo {author} {\bibfnamefont
				{A.}~\bibnamefont {Ludwig}},\ and\ \bibinfo {author} {\bibfnamefont {R.~J.}\
				\bibnamefont {Warburton}},\ }\bibfield  {title} {\bibinfo {title} {{Narrow
					optical linewidths and spin pumping on charge-tunable close-to-surface
					self-assembled quantum dots in an ultrathin diode}},\ }\href
		{https://doi.org/10.1103/PhysRevB.96.165440} {\bibfield  {journal} {\bibinfo
				{journal} {Phys. Rev. B}\ }\textbf {\bibinfo {volume} {96}},\ \bibinfo
			{pages} {165440} (\bibinfo {year} {2017})}\BibitemShut {NoStop}%
		\bibitem [{\citenamefont {Wen}\ \emph {et~al.}(1995)\citenamefont {Wen},
			\citenamefont {Lin}, \citenamefont {Jiang},\ and\ \citenamefont
			{Chen}}]{Wen1995b}%
		\BibitemOpen
		\bibfield  {author} {\bibinfo {author} {\bibfnamefont {G.~W.}\ \bibnamefont
				{Wen}}, \bibinfo {author} {\bibfnamefont {J.~Y.}\ \bibnamefont {Lin}},
			\bibinfo {author} {\bibfnamefont {H.~X.}\ \bibnamefont {Jiang}},\ and\
			\bibinfo {author} {\bibfnamefont {Z.}~\bibnamefont {Chen}},\ }\bibfield
		{title} {\bibinfo {title} {{Quantum-confined Stark effects in semiconductor
					quantum dots}},\ }\href {https://doi.org/10.1103/PhysRevB.52.5913} {\bibfield
			{journal} {\bibinfo  {journal} {Phys. Rev. B}\ }\textbf {\bibinfo {volume}
				{52}},\ \bibinfo {pages} {5913} (\bibinfo {year} {1995})}\BibitemShut
		{NoStop}%
		\bibitem [{\citenamefont {Houel}\ \emph {et~al.}(2014)\citenamefont {Houel},
			\citenamefont {Prechtel}, \citenamefont {Kuhlmann}, \citenamefont {Brunner},
			\citenamefont {Kuklewicz}, \citenamefont {Gerardot}, \citenamefont {Stoltz},
			\citenamefont {Petroff},\ and\ \citenamefont {Warburton}}]{Houel2014}%
		\BibitemOpen
		\bibfield  {author} {\bibinfo {author} {\bibfnamefont {J.}~\bibnamefont
				{Houel}}, \bibinfo {author} {\bibfnamefont {J.~H.}\ \bibnamefont {Prechtel}},
			\bibinfo {author} {\bibfnamefont {A.~V.}\ \bibnamefont {Kuhlmann}}, \bibinfo
			{author} {\bibfnamefont {D.}~\bibnamefont {Brunner}}, \bibinfo {author}
			{\bibfnamefont {C.~E.}\ \bibnamefont {Kuklewicz}}, \bibinfo {author}
			{\bibfnamefont {B.~D.}\ \bibnamefont {Gerardot}}, \bibinfo {author}
			{\bibfnamefont {N.~G.}\ \bibnamefont {Stoltz}}, \bibinfo {author}
			{\bibfnamefont {P.~M.}\ \bibnamefont {Petroff}},\ and\ \bibinfo {author}
			{\bibfnamefont {R.~J.}\ \bibnamefont {Warburton}},\ }\bibfield  {title}
		{\bibinfo {title} {{High Resolution Coherent Population Trapping on a Single
					Hole Spin in a Semiconductor Quantum Dot}},\ }\href
		{https://doi.org/10.1103/PhysRevLett.112.107401} {\bibfield  {journal}
			{\bibinfo  {journal} {Phys. Rev. Lett.}\ }\textbf {\bibinfo {volume} {112}},\
			\bibinfo {pages} {107401} (\bibinfo {year} {2014})}\BibitemShut {NoStop}%
		\bibitem [{\citenamefont {Zhai}\ \emph {et~al.}(2020)\citenamefont {Zhai},
			\citenamefont {L{\"{o}}bl}, \citenamefont {Nguyen}, \citenamefont {Ritzmann},
			\citenamefont {Javadi}, \citenamefont {Spinnler}, \citenamefont {Wieck},
			\citenamefont {Ludwig},\ and\ \citenamefont {Warburton}}]{Zhai2020a}%
		\BibitemOpen
		\bibfield  {author} {\bibinfo {author} {\bibfnamefont {L.}~\bibnamefont
				{Zhai}}, \bibinfo {author} {\bibfnamefont {M.~C.}\ \bibnamefont
				{L{\"{o}}bl}}, \bibinfo {author} {\bibfnamefont {G.~N.}\ \bibnamefont
				{Nguyen}}, \bibinfo {author} {\bibfnamefont {J.}~\bibnamefont {Ritzmann}},
			\bibinfo {author} {\bibfnamefont {A.}~\bibnamefont {Javadi}}, \bibinfo
			{author} {\bibfnamefont {C.}~\bibnamefont {Spinnler}}, \bibinfo {author}
			{\bibfnamefont {A.~D.}\ \bibnamefont {Wieck}}, \bibinfo {author}
			{\bibfnamefont {A.}~\bibnamefont {Ludwig}},\ and\ \bibinfo {author}
			{\bibfnamefont {R.~J.}\ \bibnamefont {Warburton}},\ }\bibfield  {title}
		{\bibinfo {title} {{Low-noise GaAs quantum dots for quantum photonics}},\
		}\href {https://doi.org/10.1038/s41467-020-18625-z} {\bibfield  {journal}
			{\bibinfo  {journal} {Nat. Commun.}\ }\textbf {\bibinfo {volume} {11}},\
			\bibinfo {pages} {4745} (\bibinfo {year} {2020})}\BibitemShut {NoStop}%
		\bibitem [{\citenamefont {Smith}\ \emph {et~al.}(2005)\citenamefont {Smith},
			\citenamefont {Dalgarno}, \citenamefont {Warburton}, \citenamefont {Govorov},
			\citenamefont {Karrai}, \citenamefont {Gerardot},\ and\ \citenamefont
			{Petroff}}]{Smith2005}%
		\BibitemOpen
		\bibfield  {author} {\bibinfo {author} {\bibfnamefont {J.~M.}\ \bibnamefont
				{Smith}}, \bibinfo {author} {\bibfnamefont {P.~A.}\ \bibnamefont {Dalgarno}},
			\bibinfo {author} {\bibfnamefont {R.~J.}\ \bibnamefont {Warburton}}, \bibinfo
			{author} {\bibfnamefont {A.~O.}\ \bibnamefont {Govorov}}, \bibinfo {author}
			{\bibfnamefont {K.}~\bibnamefont {Karrai}}, \bibinfo {author} {\bibfnamefont
				{B.~D.}\ \bibnamefont {Gerardot}},\ and\ \bibinfo {author} {\bibfnamefont
				{P.~M.}\ \bibnamefont {Petroff}},\ }\bibfield  {title} {\bibinfo {title}
			{{Voltage Control of the Spin Dynamics of an Exciton in a Semiconductor
					Quantum Dot}},\ }\href {https://doi.org/10.1103/PhysRevLett.94.197402}
		{\bibfield  {journal} {\bibinfo  {journal} {Phys. Rev. Lett.}\ }\textbf
			{\bibinfo {volume} {94}},\ \bibinfo {pages} {197402} (\bibinfo {year}
			{2005})}\BibitemShut {NoStop}%
		\bibitem [{\citenamefont {Dreiser}\ \emph {et~al.}(2008)\citenamefont
			{Dreiser}, \citenamefont {Atat{\"{u}}re}, \citenamefont {Galland},
			\citenamefont {M{\"{u}}ller}, \citenamefont {Badolato},\ and\ \citenamefont
			{Imamoglu}}]{Dreiser2008}%
		\BibitemOpen
		\bibfield  {author} {\bibinfo {author} {\bibfnamefont {J.}~\bibnamefont
				{Dreiser}}, \bibinfo {author} {\bibfnamefont {M.}~\bibnamefont
				{Atat{\"{u}}re}}, \bibinfo {author} {\bibfnamefont {C.}~\bibnamefont
				{Galland}}, \bibinfo {author} {\bibfnamefont {T.}~\bibnamefont
				{M{\"{u}}ller}}, \bibinfo {author} {\bibfnamefont {A.}~\bibnamefont
				{Badolato}},\ and\ \bibinfo {author} {\bibfnamefont {A.}~\bibnamefont
				{Imamoglu}},\ }\bibfield  {title} {\bibinfo {title} {{Optical investigations
					of quantum dot spin dynamics as a function of external electric and magnetic
					fields}},\ }\href {https://doi.org/10.1103/PhysRevB.77.075317} {\bibfield
			{journal} {\bibinfo  {journal} {Phys. Rev. B}\ }\textbf {\bibinfo {volume}
				{77}},\ \bibinfo {pages} {075317} (\bibinfo {year} {2008})}\BibitemShut
		{NoStop}%
		\bibitem [{\citenamefont {Latta}\ \emph {et~al.}(2009)\citenamefont {Latta},
			\citenamefont {H{\"{o}}gele}, \citenamefont {Zhao}, \citenamefont
			{Vamivakas}, \citenamefont {Maletinsky}, \citenamefont {Kroner},
			\citenamefont {Dreiser}, \citenamefont {Carusotto}, \citenamefont {Badolato},
			\citenamefont {Schuh}, \citenamefont {Wegscheider}, \citenamefont {Atature},\
			and\ \citenamefont {Imamoglu}}]{Latta2009}%
		\BibitemOpen
		\bibfield  {author} {\bibinfo {author} {\bibfnamefont {C.}~\bibnamefont
				{Latta}}, \bibinfo {author} {\bibfnamefont {A.}~\bibnamefont {H{\"{o}}gele}},
			\bibinfo {author} {\bibfnamefont {Y.}~\bibnamefont {Zhao}}, \bibinfo {author}
			{\bibfnamefont {A.~N.}\ \bibnamefont {Vamivakas}}, \bibinfo {author}
			{\bibfnamefont {P.}~\bibnamefont {Maletinsky}}, \bibinfo {author}
			{\bibfnamefont {M.}~\bibnamefont {Kroner}}, \bibinfo {author} {\bibfnamefont
				{J.}~\bibnamefont {Dreiser}}, \bibinfo {author} {\bibfnamefont
				{I.}~\bibnamefont {Carusotto}}, \bibinfo {author} {\bibfnamefont
				{A.}~\bibnamefont {Badolato}}, \bibinfo {author} {\bibfnamefont
				{D.}~\bibnamefont {Schuh}}, \bibinfo {author} {\bibfnamefont
				{W.}~\bibnamefont {Wegscheider}}, \bibinfo {author} {\bibfnamefont
				{M.}~\bibnamefont {Atature}},\ and\ \bibinfo {author} {\bibfnamefont
				{A.}~\bibnamefont {Imamoglu}},\ }\bibfield  {title} {\bibinfo {title}
			{{Confluence of resonant laser excitation and bidirectional quantum-dot
					nuclear-spin polarization}},\ }\href {https://doi.org/10.1038/nphys1363}
		{\bibfield  {journal} {\bibinfo  {journal} {Nat. Phys.}\ }\textbf {\bibinfo
				{volume} {5}},\ \bibinfo {pages} {758} (\bibinfo {year} {2009})}\BibitemShut
		{NoStop}%
		\bibitem [{\citenamefont {Reigue}\ \emph {et~al.}(2018)\citenamefont {Reigue},
			\citenamefont {Lema{\^{i}}tre}, \citenamefont {{Gomez Carbonell}},
			\citenamefont {Ulysse}, \citenamefont {Merghem}, \citenamefont {Guilet},
			\citenamefont {Hostein},\ and\ \citenamefont {Voliotis}}]{Reigue2018a}%
		\BibitemOpen
		\bibfield  {author} {\bibinfo {author} {\bibfnamefont {A.}~\bibnamefont
				{Reigue}}, \bibinfo {author} {\bibfnamefont {A.}~\bibnamefont
				{Lema{\^{i}}tre}}, \bibinfo {author} {\bibfnamefont {C.}~\bibnamefont {{Gomez
						Carbonell}}}, \bibinfo {author} {\bibfnamefont {C.}~\bibnamefont {Ulysse}},
			\bibinfo {author} {\bibfnamefont {K.}~\bibnamefont {Merghem}}, \bibinfo
			{author} {\bibfnamefont {S.}~\bibnamefont {Guilet}}, \bibinfo {author}
			{\bibfnamefont {R.}~\bibnamefont {Hostein}},\ and\ \bibinfo {author}
			{\bibfnamefont {V.}~\bibnamefont {Voliotis}},\ }\bibfield  {title} {\bibinfo
			{title} {{Resonance fluorescence revival in a voltage-controlled
					semiconductor quantum dot}},\ }\href {https://doi.org/10.1063/1.5010757}
		{\bibfield  {journal} {\bibinfo  {journal} {Appl. Phys. Lett.}\ }\textbf
			{\bibinfo {volume} {112}},\ \bibinfo {pages} {073103} (\bibinfo {year}
			{2018})}\BibitemShut {NoStop}%
		\bibitem [{\citenamefont {Vural}\ \emph {et~al.}(2020)\citenamefont {Vural},
			\citenamefont {Maisch}, \citenamefont {Gerhardt}, \citenamefont {Jetter},
			\citenamefont {Portalupi},\ and\ \citenamefont {Michler}}]{Vural2020a}%
		\BibitemOpen
		\bibfield  {author} {\bibinfo {author} {\bibfnamefont {H.}~\bibnamefont
				{Vural}}, \bibinfo {author} {\bibfnamefont {J.}~\bibnamefont {Maisch}},
			\bibinfo {author} {\bibfnamefont {I.}~\bibnamefont {Gerhardt}}, \bibinfo
			{author} {\bibfnamefont {M.}~\bibnamefont {Jetter}}, \bibinfo {author}
			{\bibfnamefont {S.~L.}\ \bibnamefont {Portalupi}},\ and\ \bibinfo {author}
			{\bibfnamefont {P.}~\bibnamefont {Michler}},\ }\bibfield  {title} {\bibinfo
			{title} {{Characterization of spectral diffusion by slow-light
					photon-correlation spectroscopy}},\ }\href
		{https://doi.org/10.1103/PhysRevB.101.161401} {\bibfield  {journal} {\bibinfo
				{journal} {Phys. Rev. B}\ }\textbf {\bibinfo {volume} {101}},\ \bibinfo
			{pages} {161401} (\bibinfo {year} {2020})}\BibitemShut {NoStop}%
		\bibitem [{\citenamefont {Zibik}\ \emph {et~al.}(2009)\citenamefont {Zibik},
			\citenamefont {Grange}, \citenamefont {Carpenter}, \citenamefont {Porter},
			\citenamefont {Ferreira}, \citenamefont {Bastard}, \citenamefont {Stehr},
			\citenamefont {Winnerl}, \citenamefont {Helm}, \citenamefont {Liu},
			\citenamefont {Skolnick},\ and\ \citenamefont {Wilson}}]{Zibik2009}%
		\BibitemOpen
		\bibfield  {author} {\bibinfo {author} {\bibfnamefont {E.~A.}\ \bibnamefont
				{Zibik}}, \bibinfo {author} {\bibfnamefont {T.}~\bibnamefont {Grange}},
			\bibinfo {author} {\bibfnamefont {B.~A.}\ \bibnamefont {Carpenter}}, \bibinfo
			{author} {\bibfnamefont {N.~E.}\ \bibnamefont {Porter}}, \bibinfo {author}
			{\bibfnamefont {R.}~\bibnamefont {Ferreira}}, \bibinfo {author}
			{\bibfnamefont {G.}~\bibnamefont {Bastard}}, \bibinfo {author} {\bibfnamefont
				{D.}~\bibnamefont {Stehr}}, \bibinfo {author} {\bibfnamefont
				{S.}~\bibnamefont {Winnerl}}, \bibinfo {author} {\bibfnamefont
				{M.}~\bibnamefont {Helm}}, \bibinfo {author} {\bibfnamefont {H.~Y.}\
				\bibnamefont {Liu}}, \bibinfo {author} {\bibfnamefont {M.~S.}\ \bibnamefont
				{Skolnick}},\ and\ \bibinfo {author} {\bibfnamefont {L.~R.}\ \bibnamefont
				{Wilson}},\ }\bibfield  {title} {\bibinfo {title} {{Long lifetimes of
					quantum-dot intersublevel transitions in the terahertz range}},\ }\href
		{https://doi.org/10.1038/nmat2511} {\bibfield  {journal} {\bibinfo  {journal}
				{Nat. Mater.}\ }\textbf {\bibinfo {volume} {8}},\ \bibinfo {pages} {803}
			(\bibinfo {year} {2009})}\BibitemShut {NoStop}%
		\bibitem [{\citenamefont {Kambs}\ and\ \citenamefont
			{Becher}(2018)}]{Kambs2018b}%
		\BibitemOpen
		\bibfield  {author} {\bibinfo {author} {\bibfnamefont {B.}~\bibnamefont
				{Kambs}}\ and\ \bibinfo {author} {\bibfnamefont {C.}~\bibnamefont {Becher}},\
		}\bibfield  {title} {\bibinfo {title} {{Limitations on the
					indistinguishability of photons from remote solid state sources}},\ }\href
		{https://doi.org/10.1088/1367-2630/aaea99} {\bibfield  {journal} {\bibinfo
				{journal} {New J. Phys.}\ }\textbf {\bibinfo {volume} {20}},\ \bibinfo
			{pages} {115003} (\bibinfo {year} {2018})}\BibitemShut {NoStop}%
		\bibitem [{\citenamefont {L{\"{o}}bl}\ \emph {et~al.}(2019)\citenamefont
			{L{\"{o}}bl}, \citenamefont {Scholz}, \citenamefont {S{\"{o}}llner},
			\citenamefont {Ritzmann}, \citenamefont {Denneulin}, \citenamefont
			{Kov{\'{a}}cs}, \citenamefont {Kardyna{\l}}, \citenamefont {Wieck},
			\citenamefont {Ludwig},\ and\ \citenamefont {Warburton}}]{Lobl2019a}%
		\BibitemOpen
		\bibfield  {author} {\bibinfo {author} {\bibfnamefont {M.~C.}\ \bibnamefont
				{L{\"{o}}bl}}, \bibinfo {author} {\bibfnamefont {S.}~\bibnamefont {Scholz}},
			\bibinfo {author} {\bibfnamefont {I.}~\bibnamefont {S{\"{o}}llner}}, \bibinfo
			{author} {\bibfnamefont {J.}~\bibnamefont {Ritzmann}}, \bibinfo {author}
			{\bibfnamefont {T.}~\bibnamefont {Denneulin}}, \bibinfo {author}
			{\bibfnamefont {A.}~\bibnamefont {Kov{\'{a}}cs}}, \bibinfo {author}
			{\bibfnamefont {B.~E.}\ \bibnamefont {Kardyna{\l}}}, \bibinfo {author}
			{\bibfnamefont {A.~D.}\ \bibnamefont {Wieck}}, \bibinfo {author}
			{\bibfnamefont {A.}~\bibnamefont {Ludwig}},\ and\ \bibinfo {author}
			{\bibfnamefont {R.~J.}\ \bibnamefont {Warburton}},\ }\bibfield  {title}
		{\bibinfo {title} {{Excitons in InGaAs quantum dots without electron wetting
					layer states}},\ }\href {https://doi.org/10.1038/s42005-019-0194-9}
		{\bibfield  {journal} {\bibinfo  {journal} {Commun. Phys.}\ }\textbf
			{\bibinfo {volume} {2}},\ \bibinfo {pages} {93} (\bibinfo {year}
			{2019})}\BibitemShut {NoStop}%
		\bibitem [{\citenamefont {Kuhlmann}\ \emph {et~al.}(2013)\citenamefont
			{Kuhlmann}, \citenamefont {Houel}, \citenamefont {Brunner}, \citenamefont
			{Ludwig}, \citenamefont {Reuter}, \citenamefont {Wieck},\ and\ \citenamefont
			{Warburton}}]{Kuhlmann2013c}%
		\BibitemOpen
		\bibfield  {author} {\bibinfo {author} {\bibfnamefont {A.~V.}\ \bibnamefont
				{Kuhlmann}}, \bibinfo {author} {\bibfnamefont {J.}~\bibnamefont {Houel}},
			\bibinfo {author} {\bibfnamefont {D.}~\bibnamefont {Brunner}}, \bibinfo
			{author} {\bibfnamefont {A.}~\bibnamefont {Ludwig}}, \bibinfo {author}
			{\bibfnamefont {D.}~\bibnamefont {Reuter}}, \bibinfo {author} {\bibfnamefont
				{A.~D.}\ \bibnamefont {Wieck}},\ and\ \bibinfo {author} {\bibfnamefont
				{R.~J.}\ \bibnamefont {Warburton}},\ }\bibfield  {title} {\bibinfo {title}
			{{A dark-field microscope for background-free detection of resonance
					fluorescence from single semiconductor quantum dots operating in a
					set-and-forget mode}},\ }\href {https://doi.org/10.1063/1.4813879} {\bibfield
			{journal} {\bibinfo  {journal} {Rev. Sci. Instrum.}\ }\textbf {\bibinfo
				{volume} {84}},\ \bibinfo {pages} {073905} (\bibinfo {year}
			{2013})}\BibitemShut {NoStop}%
	\end{thebibliography}

	%

\section{Methods}

	\textbf{Sample design.} The sample employed has been grown by combining MOVPE and MBE growth. A schematic view of the sample structure is depicted in Fig.~\ref{fig:fig1}a. Starting with MOVPE, the heterostructure was grown on a (100)-oriented n$^{+}$-doped GaAs substrate, starting with a \unit[300]{nm} thick n$^{+}$-doped GaAs layer followed by 29 n$^{+}$-doped pairs of Al$_{0.95}$Ga$_{0.05}$As (\unit[77]{nm}) and GaAs (\unit[64]{nm}) forming the bottom DBR. The MOVPE growth is completed with a last layer of Al$_{0.95}$Ga$_{0.05}$As and a thinner GaAs (\unit[31]{nm}) layer, both n$^{+}$-doped. From this point on the sample is shipped in atmospheric conditions to continue the growth with MBE. After careful oxide removal, the gate contact is formed by a \unit[28]{nm} thick GaAs (n$^{+}$, \unit[2$\times$10$^{18}$]{cm$^{-3}$}) layer and then followed by one (additional) GaAs/AlAs DBR pair. \unit[134.5]{nm} thick Al$_{0.34}$Ga$_{0.66}$As functions as the current blocking layer and enables an electrostatic potential across the optically active region. After \unit[5]{nm} of undoped GaAs functioning as spacer, the electron wetting layer state-free self-assembled InAs QDs~\cite{Lobl2019a} are grown in the Stranski-Krastanow mode with a subsequent flushing step. The resulting QD ensemble luminescence is peaked around \unit[910]{nm}. Tunnel coupling of the QDs with the Fermi sea is realized by a succeeding \unit[35]{nm} thick GaAs tunnel barrier. 
	In order to reach the back contact in a later step, an etch stop layer is added consisting of \unit[1]{nm} GaAs followed by \unit[1.5]{nm} AlAs and \unit[41]{nm} GaAs in growth direction. To further increase the extraction of photons, four more undoped GaAs/AlAs 
	DBRs are deposited. The sample is completed by an 78nm etch stop layer follwoed by the 64,25nm tick GaAs cap.\\
	The utilized device shows a QD density of $\unit[0.1-1\cdot10^{6}]{\text{cm}^{-2}}$ of QDs emitting in the wavelength range between \unit[899]{nm} and \unit[911]{nm} as measured by $\mu$-PL maps. After metallizing the highly doped substrate with a chromium-gold alloy, silver epoxy glue is used to contact the gate to the chip carrier. In order to apply a voltage to the n$^{++}$- doped top layer, UV photolithography and wet-chemical etching are used to provide access to the back contact, leaving unaffected the DBRs on top of the investigated QDs. Alternating wet-chemical etching using citric acid with hydrogen peroxide ($\text{H}_{\text{2}}\text{O}_{\text{2}}$) is employed to remove the GaAs layers, and hydrochloric acid (HCl) deluted with deionized water to remove the AlAs layers. For contacting the back contact a titanium-platinum-gold alloy is deposited via electron-beam physical vapor deposition and subsequently wire-bonded with a \unit[150]{$\mu$m} diameter In wire.\\
	
	\textbf{Experimental configuration.} Above barrier (AB) and resonant fluorescence (RF) pumping are achieved via a continuous wave diode laser and a Ti:sapphire pulsed laser system respectively. The sample is placed inside a closed-cycle He cryostat (Montana Instruments, Cryostation s50) and kept at a temperature of $T=\unit[6]{K}$. The excitation and light extraction is accomplished via a confocal microscopy setup, with the microscope objective located inside the cryostat. For RF detection, a cross-polarization setup is employed (suppression $\sim10^{7}$)~\cite{Kuhlmann2013c}. The signal is filtered utilizing a monochromator based on a transmission grating with a spectral resolution of $\sim\unit[15]{GHz}$. PCFS and TPI measurements were carried out with high efficiency $\left(\eta_{\text{SPAD}}=\unit[30]{\%}\right)$ single-photon avalanche diodes (SPAD) with a temporal resolution of $\Delta\tau_{\text{SPAD}}=\unit[350]{ps}$. The detectors utilized in the lifetime measurements had a higher temporal resolution $\left(\Delta\tau_{\text{SPAD}}=\unit[50]{ps}\right)$ at a cost of reduced efficiency $\left(\eta_{\text{SPAD}}=\unit[2]{\%}\right)$.\\
		
	\textbf{Ti-sapphire laser.} For the pulsed resonant excitation of the QDs, a Coherent Mira Ti:sapphire laser was utilized. The laser creates pulses of $\approx \unit[3]{ps}$ with a repetition rate of $ \unit[76.2]{MHz}$. For the PCFS measurements, this rate is amplified ($\times4$) to $\unit[304.8]{MHz}$.\\
	
	\textbf{Fabry-Perot interferometer and analysis.} The FPI utilized in the measurements has a free spectral range of \unit[15]{GHz} and a resolution of $\unit[100]{MHz}$ (extracted from the width of the SR function). The data in Fig.~\ref{fig:fig1}f were fit with a Voigt function convolved with the measured SR function. The Voigt profile includes homogeneous and inhomogeneous contributions: with the reasonable assumption that the homogeneous linewidth can be extracted from the decay time measurement of the trion line ($\tau_{\text{dec}}=\unit[\left(652\pm5\right)]{ps}$, yielding a homogeneous linewidth of $\Delta\nu_{\text{homo}}=\unit[\left(250\pm20\right)]{MHz}$), this contribution is kept fixed, while the inhomogeneous broadening is left as free fitting parameter.\\
	
	\textbf{Photon-correlation Fourier spectroscopy.} PCFS employs a Mach-Zehnder interferometers where both exits of the beamsplitter are collected, measured by single-photon counting modules and time correlated. In our experiment the translation stage was moved up to \unit[372]{mm} path difference, whereas every \unit[4]{mm} path difference a correlation measurement was carried out with a scanning rate of $\sim\unit[10]{fringes\,s^{-1}}$. The consequential frequency resolution in the spectral correlation is \unit[806]{MHz}. As this emitter shows close-to lifetime-limited linewidth, a Lorentzian lineshape is assumed for the analysis. Under this assumption, the frequency resolution for the measured spectrum is increased by a factor of two to \unit[403]{MHz}. The maximum spectral range covered in this framework is \unit[37.5]{GHz}. The excitation rate of $\unit[304.8]{MHz}$ would allow a temporal resolution of $\sim\unit[3]{ns}$. However, the current temporal resolution is limited to $\sim\unit[10]{ns}$, as the required small bin width at these short timescales drastically inflates the statistical error. At timescales reaching the scanning rate of the translation stage artifacts of the selfsame distort the signal. The temporal upper bound is therefore set to $\sim\unit[10]{ms}$.\\
	
	\textbf{Two-photon interference.} The experiment is performed with an unbalanced Mach-Zehnder interferometer (MZI) operating with variable time separations of $\tau_{\text{MZI}}=2,\,
	4\text{ and }\unit[9]{ns}$.

\end{document}